\documentclass{elsart}

\usepackage[dvips]{graphicx}
\usepackage{amsmath}

\usepackage{amssymb}

\newcommand{\D}{\ensuremath{\mathrm{d}}}

\newcommand{\C}{\ensuremath{\mathcal{C}}}
\newcommand{\Di}[1]{\ensuremath{\!\!\mathrm{d}#1\,}}

\newcommand{\thickone}{\mbox{$1\!\!1$}}

\newcommand{\qbinom}[2]{\ensuremath{\genfrac{[}{]}{0pt}{}{\,#1\,}{\,#2\,}_{\!q}}}

\newcommand{\bra}[1]{\ensuremath{\langle #1 |}}

\newcommand{\ket}[1]{\ensuremath{| #1 \rangle}}

\newcommand{\braket}[2]{\ensuremath{\langle #1 | #2 \rangle}}

\newcommand{\braopket}[3]{\ensuremath{\langle #1 | #2 | #3 \rangle}}

\newcommand{\basis}[1]{\ensuremath{\{\ket{#1}\}}}

\newcommand{\invq}{\ensuremath{q^{-1}}}

\newcommand{\Eff}{\ensuremath{\mathcal{F}}}

\newcommand{\ds}{\displaystyle}

\renewcommand{\Re}{\ensuremath{\mathrm{Re}}}
\renewcommand{\Im}{\ensuremath{\mathrm{Im}}}

\newcommand{\ruleoff}{\rule{\textwidth}{0.25pt}}

\newcounter{excount}
\newcounter{subexpart}
\newcounter{saveeqn} 

\newenvironment{exercise}[1]{
  \ruleoff\\
  \stepcounter{excount}\textbf{Exercise \arabic{excount}: #1}%
  \nopagebreak[4]
  \small
  \setcounter{subexpart}{0}%
  \newcommand{\subexercise}{\stepcounter{subexpart}(\alph{subexpart})\quad}%
  \setcounter{saveeqn}{\value{equation}}\setcounter{equation}{0}%
  \renewcommand{\theequation}{E\arabic{excount}.\arabic{equation}}%
}%
{
  \setcounter{equation}{\value{saveeqn}}%
  \ruleoff\\
}

\begin{document}
\begin{frontmatter}

\title{Nonequilibrium Dynamics in Low Dimensional Systems}
\author{M. R. Evans} 
\ead{m.evans@ed.ac.uk} and
\author{R. A. Blythe\thanksref{now}}
\ead{r.a.blythe@ed.ac.uk}
\address{Department of Physics and Astronomy, University of Edinburgh,
       Mayfield Road, Edinburgh EH9 3JZ, U.K. }
\thanks[now]{Present address: 
Department of Physics and Astronomy, University of Manchester,
Manchester, M13 9PL, U.K.}
       
\begin{abstract}
In these lectures we give an overview of nonequilibrium stochastic
systems.  In particular we discuss in detail two models, the
asymmetric exclusion process and a ballistic reaction model, that
illustrate many general features of nonequilibrium dynamics: for
example coarsening dynamics and nonequilibrium phase transitions. As a
secondary theme we shall show how a common mathematical structure, the
$q$-deformed harmonic oscillator algebra, serves to furnish exact
results for both systems. Thus the lectures also serve as a gentle
introduction to things $q$-deformed.
\end{abstract}

\end{frontmatter}

\section{Introduction}
In these lectures we explore the subject of nonequilibrium dynamics.
Before getting into any kind of detail let us first establish what we
mean by a nonequilibrium system.  This is best done by taking stock of
our understanding of an equilibrium system.  Consider the Canonical
(Boltzmann) distribution for a systems with configurations labelled
$\C$ each with energy $E(\C)$:
\begin{equation}
P(\C) = \frac{\exp(-\beta E(\C))}{Z}
\label{Pcan}
\end{equation}
where $\beta =1/kT$. The task is to calculate the partition function
\begin{equation}
Z = \sum_{\C} \exp(-\beta E(\C))\;,
\label{Zcan}
\end{equation}
from which all thermodynamic properties, in principle, can be
computed.  The distribution (\ref{Pcan}) applies to systems in thermal
equilibrium {\it i.e.} free to exchange energy with an environment at
temperature $T$. It can easily be generalised to systems free to
exchange particles, volume etc but it always relies on the concept of
the system being at equilibrium with its environment.

If one were interested in dynamics, for example to simulate the model
on a computer, one might choose transition rates between
configurations to satisfy
\begin{equation}
\label{db}
W(\C^\prime\to \C) {\rm e}^{-\beta E(\C^\prime)}
= W(\C\to\C^\prime) {\rm e}^{-\beta E(\C)}
\end{equation}
where $W(\C^\prime\to\C)$ is the transition rate from configuration $\C^\prime$ to
$\C$.  Condition (\ref{db}) is known as detailed balance and
guarantees (under the assumption of ergodicity to be discussed below)
that starting from some nonequilibrium initial condition the system
will eventually reach the steady state of thermal equilibrium given by
(\ref{Pcan}).  We will discuss further this dynamical relaxation
process and the properties of the steady state endowed by the detailed
balance condition in Section~\ref{Sec:Theory}.  For the moment we note
that a system relaxing to thermal equilibrium is one realisation of a
nonequilibrium system.  In recent years such relaxation dynamics have
been of special interest, for example, in the study of glassy dynamics
whereby, on timescales realisable in experiment (or simulation), the
system never reaches the equilibrium state and it is a very slowly
evolving nonequilibrium state that is observed.  This is sometimes
referred to as `off-equilibrium' dynamics.  Also let us mention the
field of domain growth whereby an initially disordered state is
quenched (reduced to a temperature below the critical temperature for
the ordered phase) and relaxes to an ordered state through a process
of coarsening of domains.  The interesting physics lies in the scaling
regime of the coarsening process which is observed before the
equilibrium (ordered) state is reached.

The other meaning of nonequilibrium refers to a system that reaches a
steady state, but not a steady state of thermal equilibrium.  Examples
of such nonequilibrium steady states are given by driven systems with
open boundaries where a mass current is driven through the system.
Thus the system is driven by its environment rather being in thermal
equilibrium with its environment.

A pragmatic definition of a nonequilibrium system that encompasses all
of the scenarios above is as a model defined by its dynamics rather
than any energy function {\it i.e.} the configurations of the model are
sampled through a local stochastic dynamics which {\it a priori} does
not have to obey detailed balance.

\subsection{Structure of these notes}
These notes broadly follow the four lectures given at the summer school.
In addition a tutorial class was held to explore
points left as exercises in the lectures. In the present notes
these exercises are included in a  self-contained form that should allow
the reader to work through them without getting stuck
or else leave them for another time and continue with the main
text. The notes are structured as follows:
in section~2 we give an overview of two simple models
that we are mainly
concerned with in these  lectures.
In section~3 we then set out the general theory of the
type of stochastic
model we are interested in and point out the technical difficulties in
calculating dynamical or even steady-state properties.  Section~4 is
an interlude in which we introduce, in a self-contained way, a
mathematical tool---the $q$-deformed harmonic oscillator---that will
prove itself of use in the final two sections.  In Section~5 we 
present the solution of the partially asymmetric exclusion process and
amongst other things how the phase diagram (Figure~\ref{fig:TASEPPD})
is generalised. In Section~\ref{Sec:SBAC} we discuss the exact
solution of a stochastic ballistic annihilation and coalescence model.

\section{Two simple models}
In this work we will focus on two exemplars of nonequilibrium systems:
the partially asymmetric exclusion process and a particle reaction
model. These models have been well studied over the years and a large
body of knowledge has been built up \cite{Privman}. We introduce the
models at this point but will come back to these models in more detail in
Sections~\ref{Sec:PASEP} and \ref{Sec:SBAC} in which we summarise some
recent analytical progress.

\subsection{Asymmetric exclusion process}
\label{Sec:ASEPintro}
\subsubsection{Model definition}
The asymmetric simple exclusion process (ASEP) is a very simple driven
lattice gas with a hard core exclusion interaction \cite{Liggett}.
Consider $M$ particles on a  one-dimensional lattice of length $N$ say.  At
each site of the lattice there is either one particle or an empty site
(to be referred to as a vacancy or hole)---there is no multiple
occupancy.

The dynamics are defined as follows: during each time interval $\Delta
T$ each particle has probability $\Delta T$ of attempting a jump to
its right and probability $q\Delta T$ of attempting a jump to its
left; a jump can only succeed if the target site is empty.

In this work will be concerned with the case $\Delta T = dt \to 0$ so
that we obtain continuous time dynamics with particles attempting hops
to the right with rate 1 and hops to the left with rate $q$. In this
limit no two particles will jump at the same time. The other limit of
$\Delta T =1$ would correspond to fully parallel dynamics where all
particles attempt hops at the same time. This type of dynamics is
employed (for $q=0$) in traffic flow modelling (see Schadschneider
this volume).

\begin{figure}
\center{
\includegraphics[width=9.6cm]{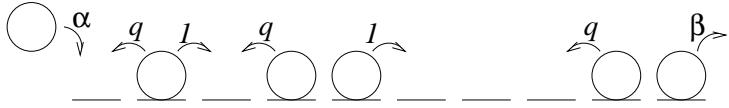}}
\caption{\label{fig:PASEP}Dynamics of the partially asymmetric
exclusion process with open boundaries.}
\end{figure}

To complete the specification of the dynamics we have to fix the
boundary conditions. It turns out these are of great significance.
The following types of  boundary conditions have been considered:
\begin{description}
\item[Periodic] 
  In this case we identify site $N+1$ with site $1$. The
  particles then hop around a ring and the number of particles is
  conserved.  
  
\item[Open boundaries] In this case particles attempt to hop into site $1$ with
  rate $\alpha$.  A particle at site $N$ leaves the lattice with rate
  $\beta$.  Thus the number of particles is not conserved at the
  boundaries.  One can also understand these boundary conditions in
  terms of a site 0 being a reservoir of particles with fixed density
  $\alpha$ and site $N+1$ being a reservoir with fixed density $1-\beta$.
  We shall primarily be concerned with these boundary conditions because,
  as first pointed out by Krug \cite{Krug91}, they can induce phase transitions.
  The dynamics is illustrated in Figure~\ref{fig:PASEP}.

\item[Closed boundaries]
  In this case the boundaries act as
reflecting walls and particles can not enter or leave the lattice.
  Thus one has a zero current of particles in the steady state
  and the system obeys detailed balance. We shall not be interested in
  this case.
  
\item[Infinite system] Finally we could consider our lattice of size
  $N$ as a finite segment of an infinite system.
\end{description}

\subsubsection{Density and current in the ASEP}
The macroscopic properties of the ASEP in which we are interested are
the steady-state current (flux of particles across the lattice) and
density profile (average occupancy of a lattice site).  These can be
expressed in terms of the binary variables $\{ \tau_i \}$, where
$\tau_i =1 $ if site $i$ is occupied by a particle and $\tau_i=0$ if
site $i$ is empty, and which together completely specify a microscopic
configuration of the system.  Then, the density at site $i$ is
defined as $\langle \tau_i(t) \rangle$ where the angular brackets
denote an average over all histories of the stochastic dynamics.
One can think of this average as being an average over an
ensemble
of systems all starting from the same initial configuration at time
0.

Let us consider for the moment the totally asymmetric dynamics $q=0$
({\it i.e.}\ no backward particle hops are permitted). One can use the
`indicator' variables $\{ \tau_i \}$ intuitively to write down an
equation for the evolution of the density.  For example, if we choose
a non-boundary site (so that we can avoid prescribing any particular
boundary conditions) we obtain
\begin{equation}
\frac{\D \langle \tau_i \rangle}{\D t}= \langle \tau_{i-1}(1 - \tau_i)
\rangle - \langle \tau_i (1 -\tau_{i+1}) \rangle \;.
\label{Ev1}
\end{equation}
Note that $\langle \tau_{i-1}(1 - \tau_i) \rangle$ gives the
probability that site $i-1$ is occupied and site $i$ is empty.  Thus,
since particles hop forward with rate 1, first term on the right hand side
gives the rate at which particles enter site $i$ and the second term
gives the rate at which they leave site $i$.

If one needs more convincing of the argument used to obtain
(\ref{Ev1}) consider what happens at site $i$ in
an infinitesimal interval $\D t$:
\begin{equation}
\label{rule}
\tau_i(t+\D t)= \left\{ \begin{array}{l@{\quad\mbox{with probability}\quad}l}
\tau_i(t) & 1 - 2 \D t \\[0.2ex]
\tau_i(t) + \left[ 1-\tau_i(t) \right] \tau_{i-1}(t)& \D t \\[0.2ex]
\tau_{i}(t)\tau_{i+1}(t) & \D t
\end{array} \right. \;.
\end{equation}
The first equation comes from the fact that with probability $1 - 2
\D t$ (dropping terms of $O(\D t^2)$), neither of the sites $i-1$ or $i$
is updated and therefore $\tau_i$ remains unchanged.  The second and
the third equations correspond to updating sites $i-1$ and $i$
respectively.  If one averages (\ref{rule}) over the events which may
occur between $t$ and $t+ \D t$ and all histories up to time $t$ one
obtains (\ref{Ev1}).

The same kind of reasoning allows one to write down straight away an
equation for the evolution of $\langle \tau_i \tau_{i+1} \rangle $.
\begin{equation}
\frac{\D \langle \tau_i \tau_{i+1} \rangle}{\D t}
 =  \langle \tau_{i-1}(1-\tau_i)\tau_{i+1} \rangle
- \langle \tau_{i}\tau_{i+1}(1-\tau_{i+2}) \rangle
\label{Ev2}
\end{equation}
or for any other correlation function $\langle \tau_i \tau_j \cdots
\rangle$.  These equations are exact and give in principle the time
evolution of any correlation function.  However, the evolution
(\ref{Ev1}) of $\langle \tau_i \rangle$ requires the knowledge of
$\langle \tau_i \tau_{i+1} \rangle$ which itself (\ref{Ev2}) requires
the knowledge of $\langle \tau_{i-1} \tau_{i+1} \rangle$ and $\langle
\tau_{i-1} \tau_i \tau_{i+1} \rangle$ so that the problem is
intrinsically an $N$-body problem in the sense that the calculation of
any correlation function requires the knowledge of all the others.
This is a situation quite common in equilibrium statistical mechanics
where, although one can write relationships between different
correlation functions, there is an infinite hierarchy of equations
which in general makes the problem intractable.

For the case of open boundary conditions the number of particles in
the system is not conserved. The evolution equations
(\ref{Ev1}-\ref{Ev2}) are then valid everywhere in the bulk but they
are modified at the boundary sites. For example (\ref{Ev1}) becomes
\begin{equation}
\frac{\D \langle \tau_1 \rangle}{\D t} = \alpha \langle (1 - \tau_1)
\rangle - \langle \tau_1 (1 -\tau_2) \rangle
\label{Ev3}
\end{equation}
\begin{equation}
\frac{\D \langle \tau_N \rangle}{\D t} = \langle \tau_{N-1}(1 - \tau_N)
\rangle - \beta \langle \tau_N \rangle\;.
\label{Ev4}
\end{equation}

In the steady state, where the time derivatives of correlation
functions are zero, (\ref{Ev1},\ref{Ev3},\ref{Ev4}) can be rewritten
as a conserved current $J$
\begin{equation}
J= \alpha \langle (1 - \tau_1) \rangle = \langle \tau_1 (1 -\tau_2)
\rangle = \cdots \langle \tau_{i-1} (1 - \tau_i)\rangle = \cdots
\langle \tau_{N-1}(1 - \tau_N) \rangle = \beta \langle \tau_N \rangle
\;.
\label{current}
\end{equation}
These equations simply express that for the density to be stationary
one must have the current into any site equal to the current out of
that site.  For example, $\alpha\langle(1-\tau_1)\rangle$ is the
probability that the leftmost site is empty multiplied by the rate at
which particles are inserted onto it, and hence yields the current of
particles entering the system.  Similarly $\beta\langle \tau_N
\rangle$ is the rate at which particles leave the system.

\subsubsection{Applications of the ASEP}
To motivate the study of this model let us consider a few
applications.  First we may consider the ASEP as a very simplistic
model of traffic flow with no overtaking (more realistic
generalisations are discussed by Schadschneider, this volume). Joining many
open boundary systems together into a network forms a very simple
model for city traffic.  Such a network appears to represent
faithfully real traffic phenomena such as jamming \cite{CLQ}.

A major motivation for the study of the ASEP is its connection to
interface dynamics and hence to the KPZ equation (noisy Burgers'
equation).  The KPZ equation is a stochastic non-linear partial
differential equation that is central to much of modern statistical
physics. In these lectures we do not have time to develop the theory
of the KPZ equation, rather we refer the reader to \cite{Krug97}. Here
we just make clear the connection to a particular interface growth
model.

\begin{figure}[b]
\center{\includegraphics[scale=0.8]{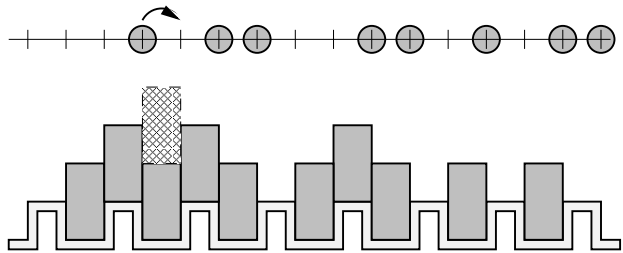}}
\caption{\label{fig:singlestep}The single-step growth model and its
mapping onto the asymmetric exclusion process with $q=0$.  Note that
the addition of a new particle to the surface (shown light-shaded)
corresponds to the rightward hop of a particle in the ASEP.}
\end{figure}

The asymmetric exclusion model may be mapped exactly onto a model of a
growing interface in $(1+1)$ dimensions \cite{MRSB} as shown in
Figure~\ref{fig:singlestep}.  The mapping is obtained by associating
to each configuration $\{\tau_i\}$ of the particles a configuration
of an interface: a particle at a site corresponds to a downwards step
of the interface height of one unit whereas a hole corresponds to an
upward step of one unit. The heights of the interface are thus defined
by
\begin{equation}
h_{i+1} - h_{i}= 1- 2 \tau_i\;.
\end{equation}
The dynamics of the asymmetric exclusion process in which a particle
at site $i-1$ may interchange position with a neighbouring hole at
site $i$, corresponds to a growth at a site $i$ which is a minimum of
the interface height {\it i.e.}  if $h_i(t)= h_{i-1}(t)-1 = h_{i+1}(t)-1$
then
\begin{equation}
h_{i} \to h_{i}(t) +2 \quad \mbox{with rate} \quad 1 \;.
\end{equation}
A growth event turns a minimum of the surface height into a maximum
thus the ASEP maps onto what is known as a single-step growth model
\cite{MRSB} meaning that the difference in heights of two neighbouring
positions on the interface is always of magnitude one unit.  Since
whenever a particle hops forward the interface height increases by two
units, the velocity $v$ at which the interface grows is related to $J$
the current in the asymmetric exclusion process by
\begin{equation}
v= 2J \;.
\label{vJ}
\end{equation}

Periodic boundary conditions for the particle problem with $M$
particles and $N-M$ holes correspond to an interface satisfying
$h_{i+N} = h_{i} + N - 2 M$, {\it i.e.} to helical boundary conditions with
an average slope $1- 2M/N$.  The case of open boundary conditions
corresponds to special growth rules at the boundaries.  Because of
this equivalence, several results obtained for the asymmetric
exclusion process can be translated into exactly computable properties
of the growing interface.

One can also go on to list further applications \cite{Schutz}. Indeed early
applications concerned biophysical problems such as single-filing
constraint in transport across membranes \cite{Heckmann} and the
kinetics of biopolymerisation \cite{MGP}.

\subsubsection{Phase diagram for $q=0$}
\begin{figure}
\center{\includegraphics[scale=0.9]{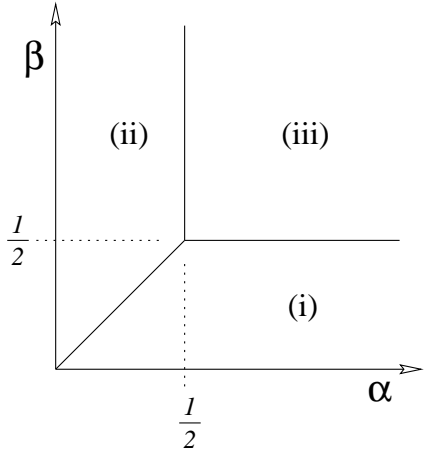}}
\caption{\label{fig:TASEPPD}Phase diagram for the totally asymmetric
exclusion process ($q=0$).}
\end{figure}

Let us now discuss some interesting results that have been derived
over the last decade for the open boundary system \cite{DEHP,SD}.  For
the moment we present them without proof, although we shall recover
some of them in the analysis of later sections.  In
Figure~\ref{fig:TASEPPD} we present the phase diagram (corresponding
to the limit $N\to\infty$) for the totally asymmetric exclusion process {\it
i.e.} when $q=0$ as predicted within a mean-field
approximation \cite{MGP,DDM}. It should be noted that we consider
the steady state in the limit $N\to \infty$, thus it is implicit
that we have taken the limit $t \to \infty$ first.
Note there are three phases---by phase it is meant a region in the
phase diagram where the current has the same analytic form; phase
boundaries divide regions where the current $J$ and bulk density
$\rho$ (defined as the mean occupancy of a site near the centre of the
lattice as $N\to\infty$) have different forms:
\begin{description}
\item[\sl (i) High density phase] For $\beta < \frac{1}{2}$ and
$\alpha > \beta $,
\begin{equation} 
J = \beta(1-\beta) \qquad \rho = 1-\beta
\label{high}
\end{equation}
\item[\sl (ii) Low density phase] For $\alpha < \frac{1}{2}$ and 
$\beta > \alpha$,
\begin{equation}
J = \alpha(1-\alpha) \qquad \rho = \alpha
\label{low}
\end{equation}
\item[\sl (iii) Maximal current phase] For $\alpha \ge \frac{1}{2}$ and 
$\beta \ge \frac{1}{2}$,
\begin{equation}
J = \frac{1}{4} \qquad \rho = \frac{1}{2}
\label{Jmax}
\end{equation}
\end{description}

Consider the high density phase. In this phase the exit rate is low
and the bulk density is controlled by the right boundary. One can
think of this as queue of traffic formed at a traffic light that
doesn't let many cars through at a time.

The low density phase is the opposite scenario where the left boundary
controls the density through a low input rate of particles. One can
think of this as cars being let onto an empty road only a few at a time.

Finally in the maximal current phase the current of particles is
saturated {\it i.e.} increasing $\alpha$ or $\beta$ any further does
not increase the current or change the bulk density.  In this phase
the density of particles decays from $\alpha$ its value at the left
boundary to $1/2$ its value in the bulk as $x^{-1/2}$ where $x$ is the
distance from the left boundary.

This phase diagram of the model has many interesting features. Firstly
we have both continuous (from high density or low density to maximal
current) and discontinuous (from low density to high density) phase
transitions even though it is a one-dimensional system.  In the former
transitions there are discontinuities in the second derivative of the
current.  In the latter although the current is continuous across the
phase transition its first derivative and the bulk density jump.
Along the transition line $\alpha=\beta <1/2$ one has coexistence
between a region of the high density phase adjacent to the right
boundary and a region of the low density phase adjacent to the left
boundary.  These results are very suggestive that the current $J$ can
be thought of as some free energy. We shall see later through an exact
solution how this can be quantified. Also note that throughout the
maximal current phase one has long-range, power-law correlations
functions {\it i.e.} long-range correlations are generic. This 
contrasts with usual behaviour
in  equilibrium systems where power-law correlations
are  seen at non-generic, critical points (an exception is
the Kosterlitz-Thouless phase that may emerge at low temperature in 
two-dimensional equilibrium systems).

\subsection{Reacting particle systems and applications}
\label{Sec:SBACintro}

We now have a look at reaction-diffusion systems wherein particles can
react with each other to form some by-product or more simply
annihilate each other.  Well-studied reaction systems are
\begin{equation}
A+A\to \emptyset \qquad  A+A\to A\;.
\end{equation}
The first reaction signifies that when two $A$ particles meet they
react and annihilate each other, the second that the two reagents
coalesce.  These systems model molecular reactions and the dynamics of
laser induced quasiparticles known as excitons \cite{exp}.  However
the particles in this type of system can also represent composite
objects such as aggregating traffic jams or, as we shall see, domain
walls in phase-ordering kinetics.

Consider first the $A+A\to \emptyset$ reaction. Clearly the steady
state of such a system is the uninteresting dead state where all
particles (except perhaps one) have been annihilated.  The true
interest lies in the late time regime wherein the system may enter a
{\em scaling} regime. By this it is meant that the system is
statistically invariant under rescaling by a typical length scale that
depends on time. In this case the length scale can be taken as the
typical distance between particles.

For diffusing $A$ particles, known results are that for $d =1$ the
typical distance between particles grows as $t^{1/2}$ or equivalently
the density of a particles decreases as $\rho \sim t^{-1/2}$.  We
shall recover this result in the next section.  The result differs
from a naive mean-field description obtained by assuming
\[ \frac{\partial \rho}{\partial t} \propto - \rho^2 \]
which would predict $\rho \sim t^{-1}$.  A similar density decay law
holds for the $A+A\to A$ process and it has been shown that the two
processes are basically equivalent \cite{Spouge,Peliti,Schutz97,DbA}.
In addition to having a range of applications, diffusive reaction
systems have served as prototypes for the development of a variety of
theoretical tools including field-theoretic techniques \cite{MG} and
the renormalisation group \cite{CT} and exact methods in one dimension
\cite{Privman,Schutz}.

One can also consider {\it ballistic} particle motion.  In this class
of models, particles move deterministically with constant velocity and
on meeting have some probability of coalescing into one particle or
annihilating.  A seminal paper by Elskens and Frisch \cite{EF} that
considered the one-dimensional
deterministic case where right moving and left moving
particles always annihilate on contact, showed the decay depends on the
initial densities of the two species.
In particular, $\rho \sim t^{-1/2}$ only if
there are equal initial densities of the right-moving and left-moving
particles.  Ballistic models apply to chemical reactions when the
inter-reactant distance is less than what would be the mean free path
of particles.  Such models may also be
applied  to various domain
growth problems \cite{Redner} and as we shall describe in section 6, the
smoothing of a growing interface \cite{KS}.

\section{Theory of Stochastic Processes}
\label{Sec:Theory}
Having now given a rough overview of some of the interest in low
dimensional nonequilibrium systems, the rest of these lectures will be
aimed at a given a deeper understanding of how these phenomena come
about in some of the simple models we have discussed.  In this section
we give a brief introduction to stochastic processes focussing on
particular aspects that will be relevant to the models to be studied
in detail in later sections.

We consider a system that can be in a finite number of microscopic
configurations and whose configuration changes according to transition
rates.  The simplest example is of a particle performing a random walk
in continuous time on a one-dimensional lattice:  the
particle hops to the right with rate $p$ and to the left with rate
$q$. (The meaning of a `rate' $W$ is that in an infinitesimal time
interval $\D t$ an event happens with probability $W\D t$.)  Since
probability is conserved we can write a continuity equation for the
probability which is referred to as `the master equation'
\begin{equation}
\frac{\partial P(x, t)}{\partial t} = p P(x-1,t) + 
q P(x+1,t) - (p+q) P(x,t)
\label{rw}
\end{equation}
The first two terms on the right hand side represent `rates in' from
other configurations ({\it i.e.} different positions of the particle);
the last term
represents the rate out of the given configuration.
Note that the master equation is linear in the probabilities. 

For the case of the random walker, the master equation (\ref{rw}) can
be solved analytically: it is a diffusion equation in discrete space
but continuous time.  If we consider a periodic lattice of $N$ sites
(site $N+1$ is identified with site 1) the preferred method is to
calculate the Fourier modes. Defining
\begin{equation}
\label{Pfourier}
\widetilde{P}(k,t)= \frac{1}{N}\sum_{x=1}^{N} z_k^x
P(x,t)\quad\mbox{with}\quad z_k = {\rm e}^{2\pi i k/N}
\end{equation}
yields from (\ref{rw})
\begin{equation}
\frac{\D \widetilde{P}(k, t)}{\D t}
= \lambda_{k} \widetilde{P}(k, t)
\quad\mbox{where}\quad
\lambda_{k} =  p(z_k-1) +q(z_k^{-1}-1)\;.
\label{lambdak}
\end{equation}
Thus
\begin{equation}
P(x,t) = \sum_{k=1}^{N} {\rm e}^{\lambda_k t}z_k^{-x}
\widetilde{P}(k,0)
\quad \mbox{with}\quad\widetilde{P}(k,0)= \frac{1}{N}
\sum_{x=1}^{N} z_k^x P(x,0)\;.
\end{equation}
The steady state corresponds to the mode $k=N$ ($\lambda =0$).  The
eigenvalues $\lambda_k$ yield decay times $\tau_k=1/\left|
\Re \lambda_k\right|$ of each mode.  For $k$ finite and $N$ large
\begin{equation}
\lambda_k \simeq (p-q) \frac{2\pi i k}{N} -(p+q)\frac{4\pi^2 k^2}{N^2}
\end{equation}
Thus the equilibration time goes like $\tau \sim N^z$ where the
dynamic exponent $z=2$ as expected of a diffusion process.


\begin{exercise}{Random walker on the 1d lattice}

\subexercise Review the solution of equation (\ref{rw}) using Fourier modes
described above.  Show that if the initial condition is $P(x,0) =
\delta_{x,x_0}$ the probability distribution at time $t$ is given by
\[
P(x,t | x_0, 0) = \frac{1}{N} \sum_{k=1}^{N} \exp \left( \lambda_k t
- \frac{2\pi i k (x-x_0)}{N} \right) \;.
\]
where $\lambda_k$ is given by (\ref{lambdak})

If you are unfamiliar with the discrete Fourier transform, you should
first show that for $1 \le k,\ell \le N$
\[
\frac{1}{N} \sum_{x=1}^{N} z_{x}^{k - \ell} = \delta_{k, \ell}
\]
and hence
\[
\tilde{f}(k) = \frac{1}{N} \sum_{x=1}^{N} z_k^{x} f(x)
\quad \Leftrightarrow \quad
f(x) = \sum_{k=1}^{N} z_{x}^{-k} \tilde{f}(k)
\]
where $z_{k}$ is as given by equation (\ref{Pfourier}).

\subexercise Consider now the same random walk problem but on an
infinite rather than periodic lattice.  Show from equation (\ref{rw})
that the generating function $F(z,t) = \sum_{x=-\infty}^{\infty}
P(x,t) (q/p)^{x/2} z^x$ obeys
\[
\frac{\partial F(z,t)}{\partial t} = \left[ -(p+q) +
\sqrt{pq}(z+z^{-1}) \right] F(z,t)
\]
and hence 
\begin{equation}
F(z,t) = F(z,0) \exp[-(p+q)t] \exp[\sqrt{pq}(z+z^{-1})t]\;.
\label{ex:rwgen}
\end{equation}
Given that the generating function of modified Bessel functions of the
first kind $I_n(x)$ is defined as
\[
\sum_{n = -\infty}^{\infty} I_n(x) z^n = \exp\left[ \frac{1}{2}
(z+z^{-1}) x \right]\;,
\]
show by expanding (\ref{ex:rwgen}) that the general solution for the
random walk problem is
\begin{equation}
\label{ex:rwsoln}
P(x,t) = \exp[-(p+q)t] \sum_{\ell = -\infty}^{\infty} \left(
\frac{p}{q} \right)^{\!\frac{x-\ell}{2}} P(\ell, 0) I_{x-\ell}(2
\sqrt{pq}\, t) \;.
\end{equation}

Some of the properties of the modified Bessel functions $I_n(x)$ can be gleaned
by considering the special case of a symmetric random walker that
begins at the origin, {\it i.e.}\ the case $p=q=\frac{1}{2}$ and $P(x,0) =
\delta_{x,0}$.  Then, $P(x,t) = \exp(-t) I_{x}(t)$.  Show then that
the property $I_n(0) = \delta_{n,0}$ immediately follows, as does the
fact that $I_{-n}(t) = I_{n}(t)$.  Also, show from (\ref{rw}) that the
Bessel functions satisfy the differential equation
\[
\frac{\D I_n(x)}{\D x} = \frac{1}{2} \left[ I_{n-1}(x) + I_{n+1}(x)
\right] \;.
\]

\end{exercise}


More generally we consider `many-body' problems. For example a system
of many particles on a lattice obeying stochastic dynamical rules.
For concreteness we consider the simple example of the one-dimensional
Ising model in which at each site $i$ there is a spin $S_i=\pm 1$
and periodic boundary conditions are imposed.

\begin{figure}[ht]
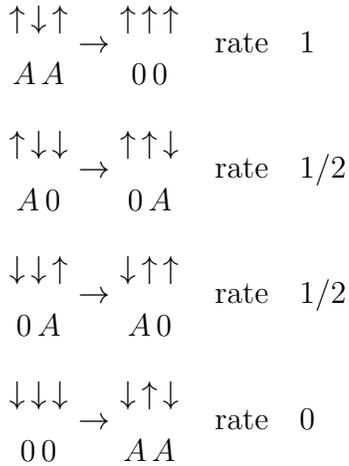

\begin{center}
\begin{eqnarray*}
\begin{array}{c}
\uparrow\, \downarrow\,  \uparrow \\
A\,A 
\end{array}
&\to& 
\begin{array}{c}
\uparrow\, \uparrow\,\uparrow\\
0\,0
\end{array}
\quad\mbox{rate}\quad 1 \\
\begin{array}{c}
\uparrow\, \downarrow\,  \downarrow\\
A\,0
\end{array} &\to& 
\begin{array}{c}
\uparrow\, \uparrow\, \downarrow\\
0\,A 
\end{array}
\quad\mbox{rate}\quad 1/2 \\
\begin{array}{c}
\downarrow\, \downarrow\,  \uparrow \\
0\,A 
\end{array}&\to& 
\begin{array}{c}
\downarrow\, \uparrow\,\uparrow\\
A\,0 
\end{array}
\quad\mbox{rate}\quad 1/2 \\
\begin{array}{c}
\downarrow\, \downarrow\,  \downarrow \\
0\,0 
\end{array}&\to& 
\begin{array}{c}
\downarrow\, \uparrow\,\downarrow\\
A\,A 
\end{array}
\quad\mbox{rate}\quad 0
\end{eqnarray*}
\caption{$T=0$ dynamics of the Ising model under Glauber dynamics.
Also indicated is the interpretation of domain walls as annihilating
particles.}
\label{IsingT=0}
\end{center}
\end{figure}

As mentioned in the introduction the dynamics of an equilibrium model
is usually chosen to satisfy detailed balance.  In the case of the
Ising model under spin-flip dynamics, whereby each spin can flip with a
certain rate according to the directions of the neighbouring spins,
the detailed balance condition on the spin-flip rates of spin $S_i$ at
site $i$ reads
\begin{equation}
\frac{W(S_i\to-S_i)}{W(-S_i\to S_i)}
= \exp\left[-2\beta (S_{i-1} + S_{i+1})S_i\right]
\label{Isingdb}
\end{equation}
(We have taken $E = - \sum_i S_i S_{i+1}$). In {\it Glauber}
dynamics (\ref{Isingdb}) is satisfied by choosing
\begin{equation}
\label{GlauberRates}
W(S_i\to-S_i)= \frac{1-S_i \tanh \left[ \beta (S_{i-1} + S_{i+1})\right] }{2}
\end{equation}
as is easily verified.  Then, in the $T= 0$ limit the spin flips for a
down-spin ($S_i=-1$) at site $i$ occur with the rates indicated in
Figure~\ref{IsingT=0} and similarly for up-spins at $T=0$.
Note the final transition where the spin flips against both its
neighbours is forbidden. Thus domains of aligned spins can not break
up.  The domain walls between aligned domains {\it i.e.} neighbouring
pairs of spins $\downarrow\,\uparrow$ or $\uparrow\,\downarrow$ make
random walks on the bonds of the original lattice and annihilate when
they meet.  As noted above, this process of diffusing `particles'
(here the domain walls) that annihilate on meeting is sometimes
denoted $A + A \to \emptyset$.  The solution 
for the mean domain length has been known for some
time \cite{TMcC,Spouge}---basically it can be solved by considering
the random walk performed by the length of a domain (distance between
neighbouring particles).

To see the solution for the mean domain size
within the kinetic Ising model \cite{Glauber}
consider the master equation which can be written
\begin{equation}
\label{KineticIsingMaster}
\frac{\partial P(\C, t)}{\partial t} = -\sum_i P(\C, t) W(S_i \to
-S_i) + \sum_i P(\C_i,t) W(-S_i \to S_i)
\end{equation}
where $\C$ is a configuration of the Ising spins and $\C_i$ is the
same configuration with spin $S_i$ flipped.  The number of domain
walls is given by $\ds \sum_i (1-S_iS_{i+1})/2$ so to study the
coarsening process, in which domain walls are eliminated
and aligned domains increase in size,
one considers the equal time correlation function
$\ds C_{k}(t)= \langle \frac{1}{N} \sum_i S_i(t) S_{i+k}(t)\rangle$
where the average indicates an average over initial conditions and
possible histories of the stochastic dynamics. It can be shown that
$C_k$ obeys a discrete diffusion equation
\begin{equation}
\frac{\partial C(k, t)}{\partial t} = -2C(k,t) +C(k+1,t) +
C(k-1,t)\quad \mbox{for}\quad k>0
\label{Ceq}
\end{equation}
with boundary condition $\quad C(0,t)=1$.

The fact that we obtain a closed set of equations for two point
correlation functions is a very happy property that allows one to
solve easily for these correlation functions---basically (\ref{Ceq})
is an equation of the same form as (\ref{rw}) but with a boundary
condition corresponding to a source of walkers at site 0: see
\cite{Bray,AF} for details.  Note that the method generalises to
finite temperature but not to higher dimensions.  Actually higher
point correlation functions are more difficult to calculate;
nevertheless it has been shown that the one-dimensional Glauber model
can be turned into a free fermion problem which then implies, in
principle at least, that any correlation function can be calculated
\cite{Felderhof}. 

Remarkably, the full domain size
probability distribution
has been explicitly calculated for
the one-dimensional $q$-state Potts model, which includes
the Ising model as the case $q=2$ \cite{DZ}. 
(Note that a calculation such
as Exercise 2 only calculates the first moment of the domain sizes).
The dynamics of domain walls in the $q$-state Potts model, 
represented as $A$ particles,
corresponds to diffusion and, on meeting, reaction according to
\begin{equation}
A+A \to \left\{ 
     \begin{array}[]{ll}
    A \quad \mbox {with probability}\quad (q-2)/(q-1)\\
    \emptyset \quad \mbox{with probability}\quad 1/(q-1)
     \end{array} 
\right.
\end{equation}
To understand this equation note
that when a domain is elimated each of the neighbouring domains
may be in one of $q-1$ states so that they
have probability $1/(q-1)$ of being in the same state
and therefore coalescing.
Thus for $q=2$ we recover $A+A\to \emptyset$ and for $q\to \infty$ 
(infinite state Potts model) 
we recover $A+A\to A$. 


\begin{exercise}{Time dependence of the 1d Ising model}

\subexercise The mean magnetisation $M(t)$ in the 1d Ising model is
defined as
\[
M(t) = \frac{1}{N} \sum_{j=1}^{N} \langle S_j \rangle \equiv
\frac{1}{N} \sum_{\{ \C \}} P(\C, t) \sum_{j=1}^{N} S_j(\C)
\]
where $N$ is the number of spins on the lattice,  $\{ \C \}$ is the
set of all $2^N$ possible spin combinations and
$S_i(\C)$ is the value of the spin at site $i$, in configuration
$\C$.
  Show for general single spin-flip dynamics as described by
(\ref{KineticIsingMaster}) that the magnetisation obeys the
differential equation
\[
\frac{\D M}{\D t} = - \frac{2}{N} \sum_{\{ \C \}} P(\C, t) \sum_{i=1}^{N} S_i
W(S_i \to - S_i)  \;.
\]

Under Glauber dynamics defined by equation (\ref{GlauberRates}) show
that
\[
\frac{\D M}{\D t} = - (1 - \tanh 2 \beta ) M(t)
\]
and hence that $M(t) = M(0) \exp(-[1-\tanh 2\beta]t)$.  Also comment
on the finite- and zero-temperature behaviour (recalling that $\beta$ is
inverse temperature).  

Hint: you will need to argue that the relation $\sum_{i} \tanh
\beta (S_{i-1} + S_{i+1}) = \sum_{i} S_i \tanh 2 \beta$ holds
for any (periodic) combination of spins.

\subexercise Show that the mean density of domain walls $n_D(t)$ in
the 1d Ising model can be expressed using the spin correlation
function $C(k,t)$ as $n_D(t) = \frac{1}{2}[1 - C(1,t)]$.  Under
zero-temperature Glauber dynamics, the evolution of $C(k,t)$ is
described by the  set of difference equations (\ref{Ceq})
which has the stationary solution $C(k) = 1 \,\forall\, k \ge 0$.
Verify this solution, and explain its physical meaning.

To obtain the time-dependence of $C(k,t)$ we introduce a function
$\epsilon(k,t)$ through $C(k,t) = 1 - \epsilon(k,t)$. 
The time-evolution
is governed by equation (\ref{rw}) with $P(k,t) = \epsilon(k,t)$,
$p=q=1$ and subject to the boundary condition $\epsilon(0,t)= 0 \,
\forall\, t$.

This boundary condition can be satisfied through an appropriate choice
of the initial conditions $\epsilon(k,0)$ in the `unphysical' region
$k<0$ (this is equivalent to using the Method of Images \cite{Feller}).  Using the general solution (\ref{ex:rwsoln}) and the
properties of the modified Bessel functions derived in exercise 1,
show that $\epsilon(0,t) = 0 \, \forall t$ if $\epsilon(-k, 0) =
\epsilon(k,0)$.  Hence show that
\[
C(k,t) = 1 - \exp(-2t) \sum_{\ell = 1}^{\infty} \left[ I_{k-\ell}(2t)
- I_{k+\ell}(2t) \right]
\]
is the solution for the spin correlation function that has $C(0,t) = 1
\,\forall\, t$ and with the initial spin orientations uncorrelated,
{\it i.e.}\ $C(k,0) = \delta_{k,0}$.  Show that this implies that for these
initial conditions, $n_D(t) = \frac{1}{2} \exp(-2t) [ I_0(2t) +
I_1(2t) ]$.

Finally, using the asymptotic form of the modified Bessel functions
\[
I_n(x) = \frac{\exp(x)}{\sqrt{2 \pi x}} \left[ 1 + \sum_{k =
1}^{\infty} \frac{(-1)^k}{k!(8x)^k} \prod_{j=1}^{k} (4n^2 - (2j-1)^2)
\right]
\]
show that the long-time behaviour of $n_D(t)$ is $n_D(t) \sim (4 \pi
t)^{-1/2} + \mathrm{O}(t^{-3/2})$.

\end{exercise}


In order to formalise stochastic processes in general we consider a
master equation
\begin{equation}
\frac{\D P(\C, t)}{\D t} = 
\sum_{\C^\prime(\neq \C)} P(\C^\prime)W(\C^\prime\to\C)
-\sum_{\C^\prime(\neq C)} P(\C)W(\C\to\C^\prime)
\end{equation}
where $W(\C\to\C^\prime)$ is the rate of transition from
configuration $\C$ to $\C^\prime$.
It is useful to combine the transition rates into a single matrix
\begin{eqnarray}
M(\C,\C^\prime) &=&W(\C^\prime\to\C)\quad\mbox{for} \quad \C^\prime\neq \C\\
M(\C,\C) &=& - \sum_{\C^\prime} W(\C\to\C^\prime)
\end{eqnarray}
so that we can write
\begin{equation}
\frac{\D P(\C, t)}{\D t} = 
\sum_{\C^\prime} M(\C, \C^\prime) P(\C^\prime, t)\; .
\end{equation}

Now let us introduce a `bra-ket' notation. Let the vector of
probabilities at time $t$ (one component for each configuration) be
$\ket{P}_t$. Thus the probability of the system being in configuration
$\C$ at time $t$ is $\langle \C | P \rangle_t$ and we can write the
master equation as
\begin{equation}
\frac{\D \ket{P}_t }{\D t} = 
 M \ket{P}_t \; .
\label{Meq}
\end{equation}
Thus the master equation looks something like a Schr\"odinger equation
with `imaginary time' and $H$ replaced by $-M$ (this is why $M$ is
sometimes referred to as a Hamiltonian).  However it should be
recalled that in quantum mechanics it is the modulus squared of the
components $\ket{P}_t$ that are probabilities whereas here the
components of $\ket{P}_t$ are probabilities.  See
\cite{ADHR,Schutz,Haye}, for example, for more details.

The formal solution of the master equation (\ref{Meq}) is
\begin{equation}
\ket{P}_t = \exp( Mt) \ket{P}_0
\label{formal}
\end{equation}
where $\ket{P}_0$ gives the initial conditions. However, a more useful
way of dealing with the problem is to consider the eigenvectors of the
transition matrix $M$. The first thing to note is that since $M$ is
not Hermitian, in general, it has different left and right
eigenvectors
\begin{eqnarray}
M\ket{\phi_\lambda} =\lambda \ket{\phi_\lambda}\label{rev}\\
\bra{\psi_\lambda} M =\lambda \bra{\psi_\lambda}\;.
\end{eqnarray}
These satisfy `bi-orthogonality'
\begin{equation}
\langle \psi_\lambda | \phi_\mu \rangle = \delta_{\lambda,\mu}\;,
\end{equation}
and in principle (assuming $M$ can be diagonalised) we may write
\begin{eqnarray}
\thickone &=& \sum_\lambda \ket{\phi_\lambda} \bra{\psi_\lambda}
\label{complete}\\
M &=& \sum_\lambda \lambda \ket{\phi_\lambda} \bra{\psi_\lambda}\;.
\end{eqnarray}
Then we have using (\ref{formal}), (\ref{complete})
and (\ref{rev})
\begin{equation}
\ket{P}_t = \sum_{n=0}^{\infty} \frac{  M^n t^n}{n!}\ \ket{P}_0
=  \sum_\lambda  \exp(\lambda t) \ket{\phi_\lambda} 
\langle \psi_\lambda | P\rangle_0 \;.
\end{equation}

Note that the real parts of all eigenvalues $\lambda$ are negative
except for one eigenvalue which is zero.  The reason for this is often
referred to as the Perron-Frobenius theorem \cite{KT}.  This theorem
holds for a matrix with non-negative entries where a sufficiently
large power of the matrix has all positive entries.  It states that
the maximum eigenvalue is non-degenerate and the corresponding left
and right eigenvectors have positive entries.

In our context one can relate the spectrum of $M$ to that of
non-negative matrix $\widetilde{M}$ obtained by adding some suitable
multiple (say minus the diagonal element with the largest magnitude) of
the identity matrix to $M$.  The assumption that a sufficiently large
power of $\widetilde{M}$ has all positive entries is equivalent to the
assumption that the (finite) system is ergodic {\it i.e.}  any
configuration can be reached from any other after sufficient time.  In
that case the system will have a unique steady state so that the zero
eigenvalue cannot be degenerate and the steady state eigenvector has
positive elements which correspond to probabilities.

In principle, all probabilities at all times can be found if one can
diagonalise the matrix $M$. Unfortunately this is only possible in a
few isolated cases. Generally this corresponds to many-particle
systems where the particles do not interact or certain `integrable'
systems that can be solved via the Bethe ansatz.

The Bethe ansatz was originally used in the context of a spin-chain
model for magnetism \cite{Bethe} and is suitable for a one-dimensional lattice
system in which the number of particles is conserved.  It is basically
a generalisation of the Fourier transform.  Let us briefly discuss it
for the case of the ASEP.  The idea is to guess the following form for
eigenvectors $\phi$
\begin{equation}
\phi(x_1, x_2, \ldots, x_M) = \sum_{Q} A(Q) z_1^{x_{Q(1)}} z_2^{x_{Q(2)}}
\cdots  z_M^{x_{Q(M)}}
\end{equation}
in which $x_i$ is the position of particle $i$ on a lattice, and $1
\le x_1 < x_2 < x_3 \ldots < x_M \le N$ with $N$ the lattice size once
again.  The sum is over all permutations $Q$ of $(1,2, \ldots, M)$ and
$A(Q)$ is some amplitude.  One uses the master equation to determine
relations that must be satisfied by $\phi(x_1, \ldots, x_M)$ in its
unphysical regions, {\it e.g.}~where two particles occupy the same site and
$x_i = x_{i+1}$ for some $i$.  After some manipulation, one arrives at a set of
nonlinear equations that must be simultaneously satisfied by all the
`momenta' $z_i$.  An early account of this method in the context of
stochastic processes is given in \cite{GS92}.  The technique is also
described in \cite{Schutz,Derrida98}.  However even for the so-called
`integrable' systems where the Bethe ansatz works it is actually very
difficult to extract the explicit eigenvectors.

Of course, it may not be possible to solve a model exactly, and in
this situation one often turns to approximation schemes.  For example,
one can try a `short time' expansion---we simply truncate the power
series defining the exponential of $M$ in (\ref{formal}) after $m$ terms
\begin{equation}
\ket{P}_t \simeq \sum_{n=0}^{m}
\frac{  M^n t^n}{n!} \ket{P}_0\;.
\end{equation}
The finite number of terms $m$ on the right hand side can be
calculated. From the series one can try to estimate, for example, the
dynamical exponent \cite{JD}.

A project less ambitious than solving the full dynamics is to
calculate the steady state probabilities (the eigenvector with
eigenvalue 0).  We denote the steady state by dropping the subscript
$t$ so that $\ket{P}$ is the eigenvector with eigenvalue zero
\begin{equation}
\frac{\D \ket{P} }{\D t} = 
 M \ket{P} =0 \;,
\label{meqn:solveme}
\end{equation}
and its components $P(\C)$ are the steady state probabilities.
Thus one has the balance conditions
\begin{equation}
\sum_{\C^\prime} W(\C\to\C^\prime) P(\C)=\sum_{\C^\prime} W(\C^\prime\to\C) P(\C^\prime) \;.
\label{SScon}
\end{equation}

Let us review some simple ways to satisfy (\ref{SScon}).  First of all
one has the case of detailed balance where
\begin{equation}
W(\C\to\C^\prime) P(\C) = W(\C^\prime\to\C) P(\C^\prime) \;.
\label{meqn:db1}
\end{equation}
However to check this condition {\it i.e.}\ to check whether detailed balance
is satisfied we first need to know the steady state!  However there is
an equivalent set of conditions called `Kolmogorov criteria' which
state that detailed balance is obeyed if and only if the transition
rates satisfy
\begin{eqnarray}
\label{meqn:db2}
\lefteqn{W(\C_{1} \to \C_{2}) W(\C_{2} \to \C_{3}) \ldots
W(\C_{n} \to \C_{1})}\hspace{1in}\nonumber \\
&& =W(\C_{1} \to \C_{n}) W(\C_{n} \to
\C_{{n-1}}) \ldots W(\C_{2} \to \C_{1})  
\end{eqnarray}
for \textit{every} possible cycle in configuration space
$\C_{1}, \C_{2},  \ldots, \C_{n}, \C_{1}$.

That (\ref{meqn:db1}) implies (\ref{meqn:db2}) is very simple to show.
To show that the converse is also true is most simply achieved
\cite{David,Kelly} by using (\ref{meqn:db1}) to determine $P(\C)$.
Then, one finds that $P(\C)$ is unique if and only if (\ref{meqn:db2})
is satisfied.

Let us consider some of the implications of the detailed balance
condition.  First note that when one has detailed balance then in the
steady state there is no net flow of probability between any two
configurations. Since there is no flow of probability there is nothing
to distinguish the forwards direction in time from backwards
direction.  Therefore running the systems backwards in time will not
change, for example, any two time correlation functions.  Thus the
system is said to be {\em reversible}.

Now consider the simple problem of a single random walker
on a periodic one-dimensional lattice, studied at the beginning
of this section.  The steady state is given by
equal probability of the walker occupying each site. However, one does
not have detailed balance as defined above unless the hop rates are
symmetric $p=q$.  But one knows that the only
effect of $p \neq q$ is to impose a drift. Clearly with a drift the
system is not reversible. But the system is reversible under
simultaneous reversal of time and parity ({\it i.e.} direction): the parity
operation results in defining the image of position $x$ as $x^\ast=
N+1-x$.  Then trivially
\begin{equation}
W(x\to x{+}1) P(x) = W(N{-}x\to N{-}x+1) P(N{-}x)
\end{equation}
since in the steady state $P(x)$ is constant in space and both
transition rates are equal to $p$.  

More generally an {\em extended}
detailed balance condition can pertain as defined in \cite{vanKampen}:
\begin{equation}
W(\C\to\C^\prime) P(\C) = W(\C^{\prime\ast}\to\C^\ast) P(\C^{\prime\ast})
\label{edb}
\end{equation}
where $\C^\ast$ is the image configuration of $\C$ under a reversible mapping. 
One requires
$(\C^\ast)^\ast= \C$ and
\begin{equation}
\sum_{\C^\prime} W(\C \to \C^\prime) = \sum_{\C^\prime} W(\C^\ast \to \C^\prime) \;.
\label{addcon}
\end{equation}
Again one can construct an equivalent condition as
follows \cite{GW}
\begin{eqnarray}
\lefteqn{W(\C_{1} \to \C_{2}) W(\C_{2} \to \C_{3}) \ldots
W(\C_{n} \to \C_{1})}\hspace{1in}\nonumber \\
&& =
W(\C_{1}^\ast \to \C_{n}^\ast) W(\C_{n}^\ast \to
\C_{{n-1}}^\ast) \ldots W(\C_{2}^\ast \to \C_{1}^\ast)  
\end{eqnarray}
for \textit{every} possible cycle in configuration space $\C_{1},
\C_{2}, \C_{3}, \ldots, \C_{n}$ along with the additional condition
(\ref{addcon}).

The condition (\ref{edb}) is known in the probabilistic literature as
{\em dynamic reversibility} \cite{Kelly}.  More recently it has
appeared in the physics literature under the name {\em pairwise
balance} \cite{SRB}. It holds for example in the
ASEP on a ring where the image configuration is obtained
by the parity operation.

\subsection{Formal solution for the steady state}
In this subsection our aim is to derive the general solution for the
steady state (\ref{meqn:solveme}) for an arbitrary transition rate
matrix $M$.  Although the solution is too general to be of any
practical use it is instructive to see how a quantity analogous to a
partition function naturally arises.

As it stands, the set of linear equations (\ref{meqn:solveme}) is
underdetermined because the determinant of $M$ is zero (one of its
eigenvalues is zero).  However, we know that a probability
distribution must be properly normalised, so consider instead the
equation
\begin{equation}
\widetilde{M}^{(i)} \ket{P}  = \ket{i}
\end{equation}
in which $\widetilde{M}^{(i)}$ is the matrix obtained from $-M$ (the
minus sign is introduced for convenience below) by replacing the
$i^\mathrm{th}$ row with all ones, {\it i.e.}~the vector $(1, 1, 1, \ldots
1)$.  The ket vector $\ket{i}$ is the basis vector corresponding to
configuration $i$.  This set of equations can now be solved and we do
so using Cramer's rule.  This states that the probability of the
system being in configuration $\C_j$ in the steady state is
\begin{equation}
\label{meqn:cramer}
P(\C_j) = \frac{\det \widetilde{M}^{(i;j)}}{\det \widetilde{M}^{(i)}}
\end{equation}
in which the matrix $\widetilde{M}^{(i;j)}$ is obtained from
$\widetilde{M}^{(i)}$ by replacing column $j$ with a vector
with a 1 at position $j$ and zeros everywhere else.  
For simplicity we choose $i=j$, and to be clear we
write out $\widetilde{M}^{(j;j)}$ explicitly.
\begin{equation}
\widetilde{M}^{(j;j)} = \left[ \begin{array}{*{7}{c}}
-M_{1,1} & \cdots & -M_{1,j-1} & 0 & -M_{1,j+1} & \cdots & -M_{1,n} \\
\vdots & \ddots & \vdots & \vdots & \vdots & \ddots & \vdots \\
-M_{j-1,1} & \cdots & -M_{j-1,j-1} & 0 & -M_{j-1,j+1} & \cdots &
-M_{j-1,n} \\
1 & \cdots & 1 & 1 & 1 & \cdots & 1 \\
-M_{j+1,1} & \cdots & -M_{j+1,j-1} & 0 & -M_{j+1,j+1} & \cdots &
-M_{j+1,n} \\
\vdots & \ddots & \vdots & \vdots & \vdots & \ddots & \vdots \\
-M_{n,1} & \cdots & -M_{n,j-1} & 0 & -M_{n,j+1} & \cdots & -M_{n,n}
\end{array} \right] \;.
\end{equation}

We note that the determinant $\det \widetilde{M}^{(j;j)}$ is equal to
the determinant of the matrix obtained from $\widetilde{M}=-M$ by
removing row $j$ and column $j$.  This determinant is called a
cofactor of $\widetilde{M}$ and will be written $f_j \equiv f(\C_j)$.
Furthermore, we observe that the determinant in the denominator of
(\ref{meqn:cramer}) is the sum of all these cofactors.  That is
\begin{equation}
Z = \sum_{j=1}^n f_j
\label{Zneq}
\end{equation}
so that
\begin{equation}
P(\C_j) = \frac{f_j}{Z} \;.
\end{equation}
We thus identify the cofactor $f_j$ with the \textit{steady-state
weight} of configuration $\C_j$ and $Z$ as a \textit{normalisation}
for the stochastic process.  In an equilibrium system, $Z$ is related
to the partition function $Z_{eq}=\sum_{\C} \exp(-\beta E(\C))$.
However, the normalisation $Z$ obtained using the above procedure is
not guaranteed to be identical to $Z_{eq}$ due to the possibility of
common factors present in all the weights $f_j$ and hence $Z$.  With
this understood, (\ref{Zneq}) extends the notion of a partition
function to nonequilibrium systems.

\subsection{Relationship of the steady-state normalisation to
eigenvalues of the transition-rate matrix}
We now note an interesting and, as far as we are aware, little-known
result that relates the normalisation $Z$ (\ref{Zneq}) to the
eigenvalues of the transition matrix.  The relation reads
\begin{equation}
\label{meqn:Zevals}
Z = \prod_{\lambda_j \neq 0} (- \lambda_j)
\end{equation}
in which the product is over
the eigenvalues of the
transition rate $M$ except the zero eigenvalue.

This result is obtained by expanding the characteristic polynomial
$\det (\lambda \thickone - M)$ in powers of $\lambda$.  One finds
\cite{LT85}
\begin{equation}
\det (\lambda \thickone - M) = (-)^n \left( \det M + (-)^{n} \lambda
\sum_{j=1}^{n} f_j + \mathrm{O}(\lambda^2) \right)
\end{equation}
with $f_j$ as defined above.  Given that $\det M = 0$ and $Z = \sum_j
f_j$ we find that
\begin{equation}
\label{meqn:Zev}
Z = \lim_{\lambda \to 0} \frac{\det (\lambda\thickone - M)}{\lambda} =
\lim_{\lambda \to 0} \frac{\prod_{j} (\lambda-\lambda_j)}{\lambda} =
\prod_{\lambda_j \neq 0} (-\lambda_j)
\end{equation}
where the last step follows because one, and only one, eigenvalue of
$M$ is zero (at least if the process is ergodic).

\subsection{Nonequilibrium phase transitions}
\label{nept}
We now consider the existence of {\it phase transitions\/} in a
stochastic process.  As in equilibrium systems, we identify phase
transitions through discontinuities in macroscopic quantities or their
derivatives.  Furthermore, at a first-order (discontinuous)
 transition we expect phase
coexistence and at a continuous transition divergences in length and
time scales.

In terms of eigenvalues of the transition matrix $M$, a phase
transition would be indicated by the clustering of one or more
eigenvalues towards zero.  For example as a discontinuous transition
is approached one would expect long-lived `excited' states since the
decay time of an eigenstate of $M$ is inversely proportional to the
real part of the corresponding eigenvalue.  These long-lived states
can be thought of as metastable states and the approach of an
eigenvalue to zero in the thermodynamic (infinite system-size) limit
suggests phase coexistence.  Near a continuous transition one has
diverging time scales (as manifested by {\it e.g.}\ a power-law decay of an
autocorrelation function) so one expects a continuous spectrum of
eigenvalues whose real parts are close to zero.

Bearing this in mind, we note that equation (\ref{meqn:Zevals})
implies that the steady-state normalisation $Z$ also approaches zero
at a phase transition in the thermodynamic limit.  Such a scenario is
analogous to Yang-Lee theory of equilibrium phase transitions, in
which it has been rigorously shown that zeros of the grand canonical
partition function in the complex-fugacity plane converge towards a
point on the positive real axis that indicates the value of the
fugacity at which a phase transition occurs.  Similar properties are
also displayed by the zeros of the canonical partition function in the
complex-temperature plane (these are also known as Fisher zeros).
Hence, it would seem likely from (\ref{meqn:Zevals}) and the above
considerations that zeros of a steady-state nonequilibrium
normalisation in the complex plane of some control parameter would
also converge to the real axis at a nonequilibrium phase transition.
This would also imply nonanalyticities in a macroscopic variable
averaged with respect to the steady-state probability distribution
function.

This idea has been pursued in the context of an asymmetric
exclusion process \cite{Arndt}.  In this case, the steady-state
normalisation was solved for its zeros in the complex-fugacity plane
and indeed they were found to approach the real axis for values of the
model parameters that corresponded to a phase transition.
Additionally, a similar approach has been tried in the context of
directed percolation \cite{DDH}, a further lattice model that exhibits a
nonequilibrium phase transition (see {\it e.g.}~\cite{Haye} for a review).

This concludes our overview of stochastic processes. We will shortly
move on to the specific models we are going to study. As mentioned in
the introduction the `$q$-deformed Harmonic Oscillator' plays a central
role in the solution of both these models.

\section{Intermezzo: What is a $q$-deformed Harmonic Oscillator?}
\label{Sec:QDHP}
First let us recall the quantum harmonic oscillator well known from
undergraduate physics. The Hamiltonian is
\begin{equation}
\hat{H} = \frac{1}{2} \hat{x}^2 + \frac{1}{2} \hat{p}^2 
\label{ho}
\end{equation}
where we have lazily set all constants $m, \hbar, \pi$ etc to unity
and where $\hat{\;}$ denotes an operator.  One way to treat the
problem is to introduce two new operators $\hat{a},\hat{a}^\dag$
through
\begin{equation}
\hat{x}= \frac{1}{\sqrt{2}}(\hat{a}+\hat{a}^\dag)
\qquad
\hat{p}= -\frac{i}{\sqrt{2}}(\hat{a}-\hat{a}^\dag)\;.
\end{equation}
Then inserting these definitions into the commutation relation
$\displaystyle \left[ \hat{x},\hat{p}\right] = i\thickone$ yields
\begin{equation}
\hat{a}\hat{a}^\dag-\hat{a}^\dagger \hat{a}= \thickone\;.
\label{boson}
\end{equation}
Also the Hamiltonian (\ref{ho}) becomes
\begin{equation}
\hat{H} = \hat{a}^\dagger \hat{a} + \frac{1}{2}\;.
\end{equation}
The operators $\hat{a},\hat{a}^\dag$ are, of course, raising and
lowering operators.  More generally they are bosonic operators, the
defining property of which is the commutation relation (\ref{boson}).
In the number (or energy) basis they obey
\begin{equation}
\hat{a} \ket{n} = \sqrt{n}\, \ket{n-1}\qquad
\hat{a}^\dagger \ket{n} = \sqrt{n+1}\, \ket{n+1}\qquad
\quad \mbox{for} \quad n\ge 0\;.
\label{energybasis}
\end{equation}
Thus in this basis $\hat{a}^\dagger \hat{a}\ket{n} =n\,\ket{n}$ and
the Hamiltonian is diagonal.  The number $n$ labels the energy
eigenstates.   The quantised energy excitations obey Bose
statistics and the operators (\ref{energybasis}) create and annihilate
bosons respectively.

The Hermite polynomials come into play when we consider the projection
of an energy eigenstate onto the position eigenbasis defined by
\begin{equation}
\hat{x} \ket{x} = x\ket{x} \;.
\end{equation}
Then
\begin{equation}
\langle x | n \rangle = {\rm e}^{-x^2} H_n(x)
\end{equation}
where $H_n(x)$ is  the Hermite polynomial of degree $n$.

We now consider a `deformation' of (\ref{boson}) by introducing a
parameter $q$  \cite{McF}
\begin{equation}
\hat{a}\hat{a}^\dag-q\hat{a}^\dagger \hat{a}= \thickone\;.
\label{qboson}
\end{equation}
Observe that
taking $q=1$ recovers bosonic operators whereas $q=-1$ gives fermionic
operators. Thus it was originally hoped that the $q$-deformed operators
might describe some new excitations interpolating between fermions and
bosons.

The operators $\hat{a}$ and $\hat{a}^\dagger$ now operate on basis
vectors $\ket{n}$ (with $n=0,1,2,\ldots$) as follows:
\begin{eqnarray}
\label{adqaction}
\hspace{2cm}\hat{a}^\dagger \ket{n} &=& \sqrt\frac{1-q^{n+1}}{1-q} \ket{n+1} \\
\hspace{2cm}\hat{a} \ket{n} &=& \sqrt\frac{1-q^n}{1-q} \ket{n-1} \;.
\label{aqaction}
\end{eqnarray}
Thus $\hat{a}^\dagger \hat{a}$, which is the analogue of the energy
operator, is diagonal in this basis
\begin{equation}
\hat{a}^\dagger\hat{a}\ket{n} = \frac{1-q^n}{1-q}\ket{n} \;.
\end{equation}
The eigenstates of the `position operator'
$\hat{x}=\hat{a}+\hat{a}^\dagger$ are given by
\begin{equation}
\hat{x} \ket{x}
= \frac{2\,x}{\sqrt{1-q}} \ket{x} \;.
\end{equation}
Finally, we identify the $q$-Hermite polynomials as
\begin{equation}
H_n(x|q)= \langle x | n \rangle \;.
\label{qHerm}
\end{equation}

\subsection{Properties of $q$-Hermite polynomials}
Using (\ref{adqaction},\ref{aqaction}) and (\ref{qHerm}) it is
straightforward to derive a recurrence relation
\begin{equation}
\label{qhreccur}
\sqrt{1-q^{n+1}}
H_{n+1}(x|q)-2xH_{n}(x|q)+ \sqrt{1-q^n}H_{n-1}(x|q)=0 \;.
\end{equation}
This can be compared to the usual recurrence relation for Hermite
polynomials
\begin{equation}
\label{hreccur}
H_{n+1}(x)-2xH_n(x)+2nH_{n-1}(x)=0  \;.
\end{equation}
To make the connection between $q$-Hermite and ordinary Hermite
polynomials we would like the recurrence relations to coincide in the
limit $q \to 1$.  The prescription is
\begin{equation}
H_{n}(x|q) \to \frac{((1-q)/2)^{n/2}}{(q;q)_n^{1/2}} H_n(\sqrt{(1-q)/2}\;x)
\end{equation}
in which both the Hermite polynomial $H_n(x|q)$ and the independent
variable $x$ are transformed.  The factors of $\sqrt{2}$ that appear
can be traced back to the fact that we set
$\hat{x}=(\hat{a}+\hat{a}^\dagger)/\sqrt{2}$ when doing the quantum
mechanics, but we take $\hat{x}=\hat{a}+\hat{a}^\dagger$ in the
$q$-deformed case.

Explicit formul\ae\ for $\braket{x}{n}$ can be found using a
generating function technique, the details of which differ slightly
depending on whether $q<1$ or $q>1$. Here we discuss in detail the
case $q<1$.

First we define a generating function $\widetilde{G}(x,\lambda)$ for
the $q$-Hermite polynomials
\begin{equation}
\label{eqn:Gdef}
\widetilde{G}(x,\lambda)=\sum_{n=0}^{\infty} \frac{\lambda^n}{\sqrt{(q;q)_n}}
\braket{x}{n} \;.
\end{equation}
where have introduced a `$q$-factorial' notation defined through
\begin{eqnarray}
\label{eqn:qfacdef}
&&(a;q)_n = \prod_{j=0}^{n-1} (1-aq^j) \\
&&(a;q)_0 = 1 \;.
\end{eqnarray}
We later encounter products of these factorials for which we use a
standard shorthand \cite{GaspRah}:
\begin{equation}
(a,b,c,\ldots;q)_n=(a;q)_n (b;q)_n (c;q)_n \ldots \;.
\label{qfacgen}
\end{equation}
The $q$-factorial in (\ref{eqn:Gdef}) is introduced for convenience as
will now become apparent.

We obtain a functional relation for $\widetilde{G}(x,\lambda)$ by
multiplying both sides of equation (\ref{qhreccur}) by
$\lambda^n/\sqrt{(q;q)_n}$ and performing the required summations:
\begin{equation}
\label{eqn:Gfuncrel}
\widetilde{G}(x,q\lambda)=(\lambda^2 - 2\lambda x + 1) \widetilde{G}(x,\lambda) \;.
\end{equation}
It is convenient to parameterise the `position' eigenstates by an
angle $\theta$ where $x=\cos \theta$ and $0\le\theta\le2\pi$,
and to replace $\widetilde{G}(x, \lambda)$ by 
$G(\theta, \lambda)$ (a function of $\theta$).
Then
(\ref{eqn:Gfuncrel}) becomes
\begin{equation}
G(\theta,\lambda)=\frac{G(\theta,q\lambda)}{(1-\lambda
e^{i\theta})(1-\lambda e^{-i\theta})}
\end{equation}
Since $q<1$ we can iterate to obtain
\begin{equation}
\label{eqn:Gnice}
G(\theta,\lambda)=\frac{1}{(\lambda e^{i\theta},\lambda
e^{-i\theta};q)_\infty}
\end{equation}
where we have used the $q$-factorial notation (\ref{qfacgen}).  Note that
$G(\theta,0)$ 
is just \braket{\theta}{0} which we are free to set to
$1$.

The infinite product $1/(x;q)_\infty$ has a well-known and easy to
verify series representation \cite{GaspRah} valid for $|x|<1$, $q<1$
\begin{equation}
\frac{1}{(x;q)_\infty}=\sum_{n=0}^{\infty} \frac{x^n}{(q;q)_n}
\end{equation}
from which, with a little effort, we may extract the form of
$\braket{\theta}{n}$. Expanding both sides of (\ref{eqn:Gnice}) in
$\lambda$ and comparing coefficients we find
\begin{equation}
\label{eqn:HP:q<1}
\braket{\theta}{n}=\frac{1}{\sqrt{(q;q)_n}} \sum_{k=0}^{n}
\qbinom{n}{k} e^{i(n-2k)\theta}
\end{equation}
where $\qbinom{n}{k}$ is the $q$-deformed binomial defined as
\begin{equation}
\label{eqn:qbinomial}
\qbinom{n}{k}= \frac{(q;q)_n}{(q;q)_{n-k} (q;q)_k}
\end{equation}
when $0 \leq k \leq n$ and zero otherwise.  In the limit $q \to 1$ the
$q$-binomial coefficient is equal to the conventional version
$\binom{n}{k}$ familiar from combinatorics.

Important properties of the $q$-Hermite polynomials are their
orthogonality and completeness.  It can be shown \cite{GaspRah} that
the set of $q$-Hermite polynomials are orthogonal with respect to a
weight function $\nu(\theta)$. That is
\begin{equation}
\label{eqn:orthog:q<1}
\int_0^\pi \Di{\theta} \braket{n}{\theta} \nu(\theta)
\braket{\theta}{m} =\delta_{n,m}
\end{equation}
where the weight function is given by
\begin{equation}
\nu(\theta)=\frac{(q,e^{2i\theta},e^{-2i\theta};q)_\infty}{2\pi}   \;.
\end{equation}
Completeness implies we can form a representation of the identity
matrix:
\begin{equation}
\label{eqn:thickone}
\int_{0}^{\pi} \Di{\theta} \ket{\theta} \nu(\theta) \bra{\theta} =
\thickone \;.
\end{equation}

It is less well known that, in a similar manner to the above
exposition, one can obtain explicit forms for $q$-Hermite polynomials
when $q>1$ \cite{Askey}. It turns out that they can be obtained from
(\ref{eqn:HP:q<1}) by making the substitution $\theta \to \pi/2-iu$
{\it i.e.} one replaces $\cos\theta$ by $i\sinh u$ and the parameter $u$
runs from $-\infty$ to $\infty$.  The weight function that
orthogonalises these polynomials is now
\begin{equation}
\label{eqn:nu:q>1}
\nu(u)=\frac{1}{\ln q}\,\frac{1}{(\invq,-\invq e^{2u},-\invq
  e^{-2u};\invq)_\infty} \;.
\end{equation}


\begin{exercise}{Coherent states, the exponential function
and their $q$-deformations}

For the `traditional' quantum mechanical raising and lowering
operators that satisfy the commutation relation (\ref{boson})
construct the vector $\ket{\mu} = \exp(\mu \hat{a}^\dagger) \ket{0}$.
Show that this is a \textit{coherent state}, {\it i.e.}\ an eigenvector of
the lowering operator $\hat{a}$ with eigenvalue $\mu$.

Now consider
a $q$-coherent state $\ket{\mu}$  satisfying the
eigenvalue equation $(1-q)^{1/2} \hat{a} \ket{\mu} = \mu \ket{\mu}$
with $\hat{a}$ and $\hat{a}^\dagger$ the $q$-deformed raising and
lowering operators that obey (\ref{qboson}).  By solving this
eigenvalue equation, show that the
$q$-coherent state $\ket{\mu}$ may be expressed as
\[
\ket{\mu} = \exp_q([1-q]^{1/2} \mu \hat{a}^\dagger) \ket{0}
\]
where the $q$-deformation of the exponential function
is defined
through
\[
\exp_q(x) = \sum_{n=0}^{\infty} \frac{x^n}{(q;q)_n} \;.
\]
Show that this  series converges always for $q>1$, for $|x|<1$
when $q<1$ and that $\lim_{q\to 1} \exp_q ([1-q]x) = \exp(x)$.

Verify also that when $|q|<1$ and $\exp_q(x)$ converges, it can also
be written as
\[
\exp_q(x) = \frac{1}{(x; q)_\infty} \;.
\]
Hint: this is most easily achieved by checking that both the series
and product representations of $\exp_q(x)$ satisfy the functional
relation $\exp_q(qx) = (1 - x) \exp_q(x)$ and that they are
numerically equal for a special value of $x=0$.

Finally, show that
\[
\braket{\theta}{\mu} = \sum_{n=0}^{\infty} \braket{\theta}{n}
\braket{n}{\mu} = G(\theta, \mu)
\]
where $G(\theta, \mu)$ is the generating function of $q$-Hermite
polynomials (\ref{eqn:Gdef},\ref{eqn:Gnice}). 

\end{exercise}


\section{Steady State Solution of Partially Asymmetric Exclusion
  Process}
\label{Sec:PASEP}
We now consider in more detail the partially asymmetric exclusion
process (PASEP) introduced in Section~\ref{Sec:ASEPintro}.  Recall
that we discussed some of the properties of the special case where the
reverse hop rate $q$ was set to zero and in particular we presented a
phase diagram for the model (Figure~\ref{fig:TASEPPD}).  We now
describe how one obtains such a phase diagram exactly for general $q$.
We do this by relating the steady-state probability distribution of
the model to products of matrices which can be mapped onto the
$q$-deformed harmonic oscillator ladder operators.

\subsection{The matrix product formulation and its quadratic algebra}
\label{sec:DEalgebra}
In this section we review the matrix approach to finding the steady
state of the model.  The approach has been the subject of lectures at
a previous summer school in this series \cite{Derrida98} and the
reader is referred there for further details.  Here the bare
essentials of the method are outlined.

Consider first a configuration of particles $\mathcal{C}$ and its
steady-state probability $P(\mathcal{C})$. We use as an ansatz for
$P(\mathcal{C})$ an ordered product of matrices $X_1 X_2 \ldots X_N$
where $X_i=D$ if site $i$ is occupied and $X_i=E$ if it is empty. To
obtain a probability (a scalar value) from this matrix product, we
employ two vectors $\bra{W}$ and $\ket{V}$ in the following way:
\begin{equation}
\label{eqn:Pansatz}
P(\mathcal{C})=\frac{\braopket{W}{X_1 X_2 \ldots X_N}{V}}{Z_N} \;.
\end{equation}
Note that here the matrices, bra and ket vectors are in an auxiliary
space and are not related to the space of configurations and the
probability ket vector we considered earlier.  The factor $Z_N$ is
included to ensure that $P(\mathcal{C})$ is properly normalised. This
latter quantity, analogous to a partition function as was discussed in
Section~\ref{Sec:Theory}, has the following simple matrix expression
through which a new matrix $C=D{+}E$ is defined:
\begin{equation}
\label{eqn:Zdef}
Z_N=\braopket{W}{(D+E)^N}{V}=\braopket{W}{C^N}{V} \;.
\end{equation}

Note that if $D$ and $E$ do not commute $P(\mathcal{C})$ is a function
of both the number and position of particles on the lattice, as
expected for a non-trivial steady state. The algebraic properties of
the matrices can be deduced from the master equation for the process
\cite{DEHP}. It can be shown that sufficient conditions for equation
(\ref{eqn:Pansatz}) to hold are
\begin{eqnarray}
\label{DEcom}
DE-qED &=& D+E \\
\label{EW}
\alpha \bra{W}E &=& \bra{W} \\
\label{DV}
\beta D \ket{V} &=& \ket{V} \;.
\end{eqnarray}

In this formulation steady state correlation functions are easily
expressed.  For example, the mean occupation number (density) of site
$i$ may be written as
\begin{equation}
\label{eqn:taudef}
\langle \tau_i \rangle = \frac{\braopket{W}{C^{i-1}DC^{N-i}}{V}}{Z_N} \;.
\end{equation}

To get a feel for why the conditions (\ref{DEcom}--\ref{DV}) give the
correct steady state consider the current $J$ of particles between
sites $i$ and $i+1$.
\begin{equation}
J \equiv \langle \tau_i (1-\tau_{i+1})\rangle
-q\langle (1-\tau_i)\tau_{i+1}\rangle \label{jdef}
\end{equation}
In the matrix product formulation the current becomes
\begin{equation}
J = \frac{\braopket{W}{C^{i-1}(DE-qED)C^{N-i-1}}{V}}{Z_N} \;.
\end{equation}
Now using (\ref{DEcom}) and the fact that $C=D+E$ we find
\begin{equation}
J = \frac{\braopket{W}{C^{N-1}}{V}}{Z_N} = \frac{Z_{N-1}}{Z_N} \;.
\end{equation}
We see that, as is required in the steady state, the current is
independent of the bond chosen.

To see that the current into site one reduces to the same expression
we use (\ref{EW})
\begin{eqnarray}
J =  \alpha \langle 1-\tau_1\rangle
= \frac{\alpha \langle W | EC^{N-1}|V \rangle}{Z_N}
= \frac{Z_{N-1}}{Z_N}
\label{JZ}
\end{eqnarray}
Also the current out of site $N$ reduces to the same expression
using (\ref{DV})
\begin{equation}
J =  \beta \langle \tau_N \rangle
= \frac{\beta \braopket{W}{C^{N-1}D}{V}}{Z_N}
= \frac{Z_{N-1}}{Z_N}
\end{equation}
Note that the fact that all the above expressions for the current are
equivalent is a necessary condition for the matrix formulation to be
correct. A sufficient condition is to check that the master equation
(\ref{meqn:solveme}) is satisfied for all $2^N$ configurations.  The
algebra (\ref{DEcom}--\ref{DV}) allows one to do this systematically
\cite{DEHP,DE97}.

Our task now is to evaluate the matrix products in the above
expressions for $Z_N$, $J$ and $\tau_i$ by applying the rules
(\ref{DEcom}--\ref{DV}).

In \cite{DEHP} the case $q=0$ was treated by using (\ref{DEcom})
repeatedly to `normal-order' matrix products: that is, to obtain an
equivalent sum of products in which all $E$ matrices appear to the
left of any $D$ matrices.  For example, consider powers of $C$
\begin{eqnarray*}
C^0 &=&1\\
C^1 &=&D+E\\
C^2 &=&D^2+E^2 +(1+q)ED + D+E\\
&\vdots&\\
C^N &=& \sum_{n=0}^{N}\sum_{m=0}^{N-n}a_{N,n,m} E^nD^m
\end{eqnarray*}
Then finding a scalar value  would be
straightforward using (\ref{EW}) and (\ref{DV}):
\begin{equation}
\label{eqn:canonicalform}
Z_N = \braopket{W}{C^N}{V} = \sum_{n,m}a_{N,n,m} \alpha^{-n} \beta^{-m}
\;.
\end{equation}
The difficulty with this approach lies in the combinatorial problem of
finding the coefficients $a_{N,n,m}$.  This was solved for $q=0$ in
\cite{DEHP}. However, the solution 
(and the problem of actually calculating the current and
correlation functions) for arbitrary $q$ 
remained open for some time although the phase
diagram was conjectured \cite{Sandow}.  
Recently the problem
has been overcome by making the connection with the $q$-Hermite
polynomials discussed in Section~\ref{Sec:QDHP} \cite{Sasamoto1,BECE}.

\subsection{Connection with the $q$-deformed harmonic oscillator}
\label{sec:qalgebra}
To make the connection with the $q$-deformed harmonic oscillator, let us
define
\begin{eqnarray}
\label{eqn:Dofa}
D &=& \frac{1}{1-q}+\frac{1}{\sqrt{1-q}}\:\hat{a} \\
\label{eqn:Eofa}
E &=& \frac{1}{1-q}+\frac{1}{\sqrt{1-q}}\:\hat{a}^\dagger
\end{eqnarray}
One finds that the algebraic rules (\ref{DEcom}--\ref{DV}) reduce
to
\begin{eqnarray}
\hat{a}\hat{a}^\dagger&-&q\hat{a}^\dagger \hat{a} = \thickone 
\label{pasepqho1}\\
\bra{W} \hat{a}^\dagger &=& 
\frac{w}{\sqrt{1-q}}\bra{W}\quad\mbox{where}\quad
w=\frac{1-q}{\alpha}-1 
\label{pasepqho2}\\ 
\hat{a} \ket{V} &=& \frac{v}{\sqrt{1-q}} \ket{V}
\quad\mbox{where}\quad
v=\frac{1-q}{\beta}-1  \;.
\label{pasepqho3}
\end{eqnarray}
Thus $D$ and $E$ 
are related to the $q$-bosonic operators discussed in
Section~\ref{Sec:QDHP}; such a relationship was first pointed out in
\cite{Sandow}. Also we see that
$\bra{W}$ and $\ket{V}$ are eigenvectors of the $q$-bosonic
operators. Such eigenvectors are known as coherent states (see exercise 3).
In the
oscillator's ``energy'' eigenbasis \basis{n} (\ref{energybasis}) one
can check that they have components
\begin{equation}
\label{eqn:VWofn}
\braket{n}{V}=\frac{v^n}{\sqrt{(q;q)_n}}
\qquad
\braket{W}{n}=\frac{w^n}{\sqrt{(q;q)_n}}\;.
\end{equation}
For the moment we consider $v<1$, $w<1$ so that the vectors
(\ref{eqn:VWofn}) are normalisable.

We introduced earlier a matrix $C$ which appears in the expressions
for the mean particle density and current. We now see that this matrix
can be written as a linear combination of the identity $\thickone$ and
the ``position'' operator $\hat{x}=\hat{a}+\hat{a}^\dagger$:
\begin{equation}
\label{eqn:Cofx}
C = D+E = \frac{2}{1-q}\:\thickone + \frac{1}{\sqrt{1-q}}\:\hat{x}
\;.
\end{equation}
As we saw in Section~\ref{Sec:QDHP} the eigenstates of the oscillator
in the co-ordinate representation are continuous $q$-Hermite
polynomials.  Clearly, the eigenvectors of $C$ are the same as those
for $\hat{x}$ and therefore knowledge of them permits diagonalisation
of $C$.

Let us illustrate how to apply the procedure to obtain an expression
for the normalisation $Z_N$ when $q<1$. First we insert a complete set
of states into the expression for the normalisation (\ref{eqn:Zdef}):
\begin{equation}
Z_N=\int_0^\pi \Di{\theta} \nu(\theta)
\braopket{W}{C^N}{\theta}\braket{\theta}{V} \;.
\end{equation}
By design, the matrix $C$ is acting on its eigenvectors, so using
(\ref{eqn:Cofx}) and 
\begin{equation}
\hat{x} \ket{\theta} = \frac{2\,\cos \theta}{\sqrt{1-q}} \ket{\theta}
\end{equation}
we obtain
\begin{equation}
\label{eqn:Zint}
Z_N=\int_{0}^{\pi} \Di{\theta} \nu(\theta) \braket{W}{\theta} \left(
\frac{2(\cos\theta+1)}{1-q} \right)^{\!\!N} \braket{\theta}{V} \;.
\end{equation}

Now we know the form of $\bra{W}$ and $\ket{V}$ in the $\basis{n}$
basis (\ref{eqn:VWofn}).
Therefore we insert a complete set of the basis vectors in
(\ref{eqn:Zint}) to find
\begin{equation}
\braket{\theta}{V}=\sum_{n=0}^{\infty} \braket{\theta}{n}
\braket{n}{V} = \sum_{n=0}^{\infty} \frac{v^n}{\sqrt{(q;q)_n}}
\braket{\theta}{n} \;.
\end{equation}
Observe that the final sum in this equation is nothing but
(\ref{eqn:Gdef}), the generating function of the $q$-Hermite
polynomials! Thus, when $|v|<1$ and $|w|<1$, we may write
\begin{equation}
\braket{\theta}{V}=G(\theta,v)\qquad
\braket{W}{\theta}=G(\theta,w)
\end{equation}
where $G(\theta,\lambda)$ is given by expression (\ref{eqn:Gnice}).

Putting all this together, we arrive at an exact integral form for the
normalisation
\begin{equation}
\label{eqn:Zint:q<1:short}
Z_N = \left( \frac{1}{1-q} \right)^{\!\!N}
\int_0^\pi \Di{\theta} \nu(\theta) [2(1+\cos\theta)]^N G(\theta,w)
G(\theta,v) 
\end{equation}
which, written out more fully and using the notation (\ref{qfacgen}),
reads
\begin{equation}
\label{eqn:Zint:q<1:long}
Z_N = \frac{(q;q)_\infty}{2\pi} \left( \frac{1}{1-q} \right)^{\!\!N}
\int_0^\pi \Di{\theta} [2(1+\cos\theta)]^N
\frac{(e^{2i\theta},e^{-2i\theta};q)_\infty}
{(ve^{i\theta},ve^{-i\theta},we^{i\theta},we^{-i\theta};q)_\infty} \;.
\end{equation}

When $|v| >1$ or $|w|>1$ equation (\ref{eqn:Zint:q<1:short}) is not
well-defined because $G(\theta,\lambda)$ does not converge when
$|\lambda|>1$.  Rather than finding a representation of the quadratic
algebra (\ref{DEcom}--\ref{DV}) that does not suffer from this
problem, one can simply analytically continue the integral
(\ref{eqn:Zint:q<1:long}) to obtain $Z_N$ when $|v|$ or $|w|$ takes on
a value greater than one.  Since a contour integral is defined by the
residues it contains, to analytically continue the integral one simply
has to follow the poles of the integrand as they move in and out of
the original integration contour (see exercise 4).

One can also apply the procedure of the previous section to find an
integral representation of the normalisation for the case of $q>1$.
The only difference is that we must use the $q>1$ polynomials.  It
turns out that this amounts to substituting $\cos\theta$ by $i\sinh u$
with the limits on $u$ now running from $-\infty$ to $+\infty$  and
using the weight function (\ref{eqn:nu:q>1})
\begin{equation}
\label{eqn:Zint:q>1:short}
Z_N = \left( \frac{1}{1-q} \right)^{\!\!N} 
\int_{-\infty}^{\infty} \Di{u} \nu(u) [2(1+i\sinh u)]^N G(u,w) G(u,v)
\;.
\end{equation}

It is possible to derive an alternative expression for $Z_N$ which
takes the form of a finite sum rather than an integral and is valid
for {\em all} values of the model parameters. Such an expression
implies the solution of the combinatorial problem of reordering
operators $D,E$ under the rule (\ref{DEcom}).  In \cite{BECE} the
formula (eq. 55 in that paper)
was derived by explicitly evaluating the integral
(\ref{eqn:Zint:q<1:short}) using further properties of $q$-Hermite
polynomials and their generating functions. 

\subsection{Phase diagram}
The phase diagram for the partially asymmetric exclusion process is
now obtained by calculating the current through the relation
$J=Z_{N-1}/Z_N$ and the above exact expressions for the normalisation
$Z_N$.

\begin{table}[b]
\begin{center}
\begin{tabular}{c|c|c}
Phase &Region & Current $J$\\[0.5ex] \hline & \\ 
(i) & $\beta\le\frac{1-q}{2}$, $\alpha\ge\beta$ & 
$\displaystyle \frac{\beta(1-q-\beta)}{1-q}$\\[3ex]
(ii) & $\alpha\le\frac{1-q}{2}$, $\beta\ge\alpha$ & $\displaystyle
\frac{\alpha(1-q-\alpha)}{1-q}$\\[3ex] 
(iii) & $\alpha\ge\frac{1-q}{2}$, $\beta\ge\frac{1-q}{2}$ & 
$\displaystyle \frac{1-q}{4}$\\[3ex]
\end{tabular}
\end{center}
\caption{\label{table:forward}The $N\to\infty$ forms of the particle
  current in the forward biased phases.}
\end{table}

\begin{figure}[htb]
\begin{center}
\includegraphics[scale=1]{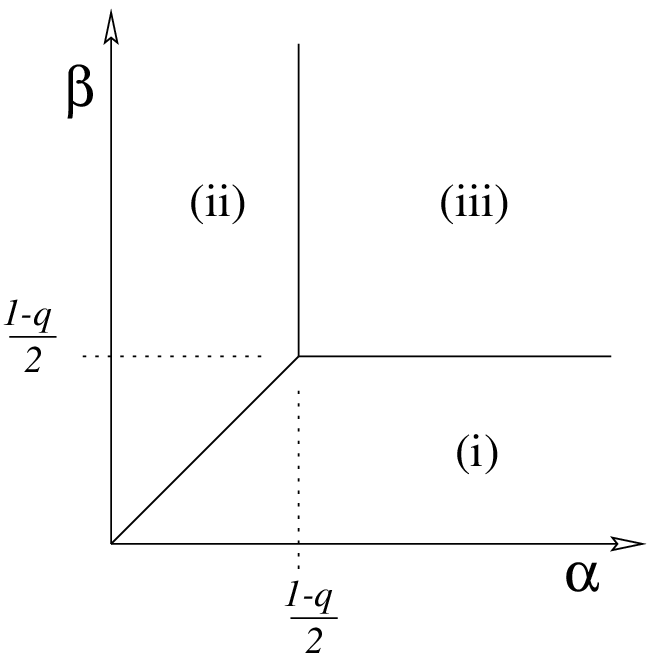}
\end{center}
\caption{\label{fig:PASEPPD}The phase diagram for the current in
the forward-bias regime of the PASEP.}
\end{figure}

For the case of {\it forward bias\/} ($q<1$) in which particle hops
are biased from the left boundary (where particles are inserted) to
the right boundary (where they are removed) one expects a nonvanishing
current in the thermodynamic limit $N \to \infty$.  This is
calculated by applying the saddle-point method to the integral
(\ref{eqn:Zint:q<1:long})---see exercise 4.  
It turns out that for large $N$
\begin{equation}
\braopket{W}{C^N}{V} \sim a^N N^{-\gamma}
\end{equation}
Then in the thermodynamic limit $J \to 1/a$.
Phase transitions arise from
different asymptotic forms of the current
due to the 
analytic
continuation (see exercise 4) for the integral
(\ref{eqn:Zint:q<1:long}).
Ultimately one finds the expressions presented in
Table~\ref{table:forward} and thence the phase diagram shown in
Figure~\ref{fig:PASEPPD}.

Note that the current as given by (\ref{JZ}) is a ratio
of two partition functions of size $N-1$ and $N$.
In equilibrium statistical mechanics (in the large $N$ limit)
this ratio is equivalent to the fugacity $z$
(within the grand canonical ensemble).
Recalling that $z= {\rm e}^\mu$ and the chemical potential
$\mu$ is equivalent to the Gibbs free energy per particle,
gives credence to our claim in section 2.1.4 that the current $J$ acts
as a free energy for the system.

We note that this phase diagram has a very similar form to that of
Figure~\ref{fig:TASEPPD} which applied for the special case of total
asymmetry $q=0$.  That is, one finds the same three phases, namely a
(i) high density, (ii) low density and (iii) maximal current phase.
As discussed in Section~\ref{Sec:ASEPintro} the transition between the
low and high density phases (i) and (ii) is first order and those to
the maximal current phase (iii) are second order.  Note from
Table~\ref{table:forward} that, for example, the 
first derivative of the
current is discontinuous at the transition between the low and high
density phases.

Similar methods can be used to find the density profiles in the PASEP:
the details of the calculations and results are presented in
\cite{Sasamoto2}.  
It turns out
that the low- and high-density phases subdivide into
three subphases for nonzero $q$ less than one
\cite{Sasamoto2}.  In each of the low
density subphases, the bulk density is $\rho=\alpha/(1-q)$; however
what distinguishes the subphases is that the decay of the density
profile from the right boundary takes a different form in each.  The
same is true of high-density phase, except that $\rho=(1-\beta)/(1-q)$
and that the density decay is at the left boundary.


\begin{exercise}{Evaluation of the current in the PASEP with $q<1$}

Show that the integral representation of the normalisation for $q<1$,
equation (\ref{eqn:Zint:q<1:long}), can be recast as a contour
integral
\begin{equation}
\label{ex:oint}
Z_N = \oint_C \D z \exp(N h(z)) g(z)
\end{equation}
where $C$ is the unit circle centred on the origin,
\[
h(z) = \ln(2 + z + z^{-1}) - \ln (1-q) \quad \mbox{and} \quad
g(z) = \frac{1}{4\pi iz}
\frac{(q,z^2,z^{-2};q)_\infty}{(vz,v/z,wz,w/z;q)_\infty} \;.
\]
The large number of singularities in the integrand at the origin makes
this a difficult integral to evaluate exactly using, {\it e.g.}, the residue
theorem.  Instead, we shall use this contour integral to obtain the
currents in the PASEP for large system size.

First, we will apply the saddle-point method to the above integral
which is valid for the range $|v|<1, |w|<1$.  The idea is to deform
the contour of integration such that it passes through a saddle point
in $\Re[h(z)]$ along the path of steepest descent (this is also the
line where $\Im[h(z)] = \mbox{const}$).  Show that this saddle point is
at $z_0=1$ and that the path of steepest descent (at least near $z_0$)
is parallel to the imaginary axis.  Also, show that $g(z_0)=g^\prime(z_0)=0$.

Insert Taylor expansions of $h(z)$ and $g(z)$ (to second order) about
the saddle point $z_0$ into (\ref{ex:oint}) to obtain
\[
Z_N = \exp(N h(z_0)) \oint_C \D z \exp \left( -\frac{1}{2} N
h^{\prime\prime}(z_0) (z-z_0)^2 \right)
\frac{g^{\prime\prime}(z_0)}{2} (z-z_0)^2 \;.
\]
You should now convince yourself that the deformation of the
integration contour to pass through $z_0$ along the path of steepest
descent in $\Re[h(z)]$ is equivalent to making the substitution $z =
z_0 + i x$ with $x$ running from $-\infty$ to $+\infty$.  Hence show
that the resulting Gaussian integral will give rise to an expression
of the form
\[
Z_N \sim K_1 \exp(N h(z_0)) N^{-3/2} \;.
\]
Note: to save time, do not evaluate the constant $K_1$ (although you
may like to check that it is real and positive).

Now show that the current in the PASEP for $|v|<1$, $|w|<1$ and
large $N$ behaves as
\[
J = \frac{Z_{N-1}}{Z_N} \sim \frac{1}{\exp h(z_0)} = \frac{1-q}{4} \;.
\]

There now remains the question of what to do when one of the control
parameters, say $v$, has a magnitude greater than $1$.  When $|v|<1$,
the contour $C$ encloses the poles of $g(z)$ at $z = v, q v, q^2 v,
\ldots$ but excludes those at $z = 1/v, q/v, q^2/v, \ldots$.  As $v$
is increased, the pole at $z=v$ moves outside $C$ and that at $z=1/v$
moves inside.  To obtain the analytic continuation of $Z_N$, one must
add to the saddle-point expression for the integral over $C$ above the
residue at $z=v$ and subtract that at $z=1/v$.  Show that this yields
\[
Z_N \sim K_1 \exp(N h(z_0)) N^{-3/2} + K_2 \exp(N h(v)) \;.
\]
Again try to avoid explicitly calculating the positive constant $K_2$.
Verify that $h(v) > h(z_0)$ and thus that, for sufficiently large $N$,
$Z_N \sim K_2 \exp(N h(v))$.  Hence show that for $1 < v < 1/q$
\[
J = \frac{Z_{N-1}}{Z_N} \sim \frac{1}{\exp h(v)} = \frac{\beta
(1-q-\beta)}{1-q} \;.
\]
Now convince yourself that as $v$ is increased yet further, the
contribution from the additional poles that have entered/exited the
unit circle is smaller than that from those at $z=v,1/v$.  Consider
also the effect of increasing $w$ to a value less than $v$ and show
that when $w>v, w>1$
\[
J \sim \frac{1}{\exp h(w)} = \frac{\alpha (1-q-\alpha)}{1-q} \;.
\]
Finally, use the information learned from this analysis to construct
the phase diagram (Figure 5) for the current in the $\alpha$-$\beta$ plane.

\end{exercise}


For $q>1$ one has a reverse bias regime \cite{BECE} where the boundary
conditions force a current through the system against the direction of
the reverse bias.  This gives rise to a new phase in which the current
decreases exponentially with the length of the system as $J \sim
q^{-N/2}$.  This phase is of interest in the context of `backbend
dynamics' \cite{TB,RB} where, for example, a fluid in a permeable
medium has to traverse a pore oriented against the direction of
gravity.

\begin{figure}[t]
\begin{center}
\includegraphics[scale=0.8]{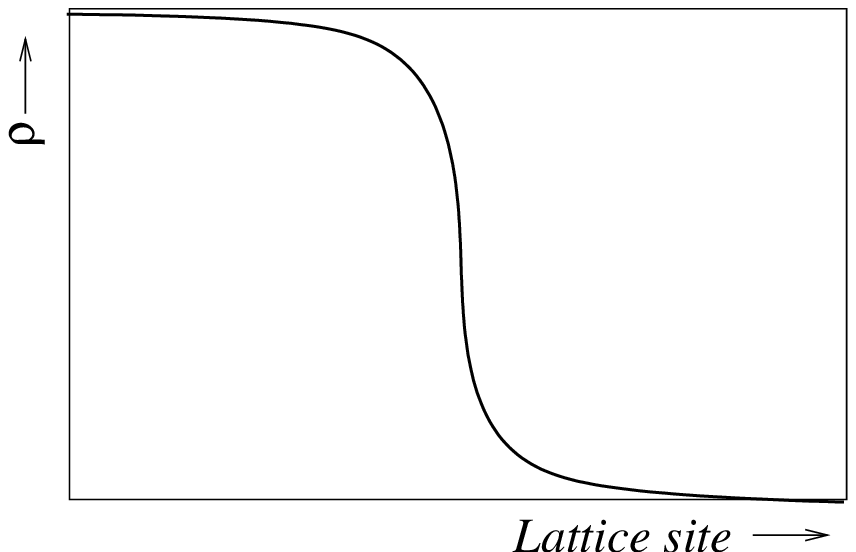}
\end{center}
\caption{\label{fig:sigmoid}Proposed form of the density profile in
the reverse-bias regime of the PASEP.}
\end{figure}

One can predict the
form of the density profile in the reverse bias regime
as follows.  We expect a high-density
region at the left edge of the lattice, towards which the particles
are biased but cannot escape, and a low-density region at the right
edge of the lattice.
Thus the density profile is a Fermi-like function
as illustrated in
Figure~\ref{fig:sigmoid}.  
In order that the profile is stationary
the typical time for
a particle to diffuse (against the bias)
from the edge of the
high-density region to the right boundary
must be equal to the typical
time of  a hole to diffuse
from the edge of the low-density region
to the  left boundary.  Since the diffusion rates of both
particles and holes are the same,
the boundary between the high- and
low-density regions must be in the centre of the lattice, as
illustrated in Figure~\ref{fig:sigmoid}.

\subsection{Generalisations}
Here we give a brief overview of other related exclusion processes.
Firstly the features of the phase diagram for $q<1$ (Figure 5) appear
robust for stochastic one-dimensional driven systems 
and can be predicted from considerations of domain wall dynamics
\cite{KSKS,HJPS}.
This has been confirmed through the exact solution of the open
boundary system with fully parallel dynamics \cite{ERS,dGN}.  However
when the bulk dynamics become deterministic the maximal current phase
is lost.

For the case of symmetric exclusion $q=1$ (no bulk drive) one does not
have phase transitions; the current is easy to calculate and decays
linearly with system size \cite{Spohn,SS,SMW}.  Intuitively one can
understand this as diffusion between two particle reservoirs of
different density. The diffusion current will be proportional to the
concentration gradient which decreases as $1/N$.  Thus although the
bulk dynamics is symmetric the boundary conditions force a weak
current of particle through the system.

For $q=1$ a recent work \cite{DLS01} has determined the probability of
observing a particular coarse-grained density profile $\rho(x)$.  In
an equilibrium system, this probability is given by
$\exp(-N\Eff[\rho])$ in which $\Eff[\rho]$ is a local functional that
expresses the free energy difference between the equilibrium profile
and $\rho$.  From the matrix product approach
a form for a functional that plays a similar role in the
symmetric exclusion process was found in \cite{DLS01} and is shown to
be \textit{nonlocal}.  For the particular boundary conditions where
the dynamics satisfy detailed balance, $\Eff[\rho]$ reduces to the
free energy functional known from equilibrium statistical physics and
so the nonlocal functional would seem to generalise a well-known
equilibrium concept.

An interesting kind of boundary-induced phase transition, manifesting
spontaneous symmetry breaking, is found when the asymmetric exclusion
process is generalised
to two oppositely moving species of particle: one species is injected
at the left, moves rightwards and exits at the right; the other
species is injected at the right, moves leftwards and exits at the
left \cite{EFGMPRL}.  Intuitively one can picture the system as a narrow
road bridge: cars moving in opposite directions can pass each other
but cannot occupy the same space.  The model has a left-right symmetry
when the injection rates and exit rates for the two species of
particles are symmetric.  However for low exit rates ($\beta$) this
symmetry is broken and the lattice is dominated by one of the species
at any given time. This implies that the short time averages of
currents and bulk densities of the two species of particles are no
longer equal.  Over longer times the system flips between the two
symmetry-related states.  In the $\beta \to 0$ limit the mean flip
time between the two states has been calculated analytically and shown
to diverge exponentially with system size \cite{GLEMSS}.  Thus the
`bridge' model provided a first example of spontaneous symmetry
breaking in a one dimensional system.

In the models discussed so far 
the open boundaries can be thought of as
inhomogeneities where the order parameter (particle density) is not
conserved.  Inhomogeneities which conserve the order parameter can be
considered on a periodic system. Indeed a single defect bond on the
lattice (through which particles hop more slowly) is sufficient to
cause the system to separate into two macroscopic regions of different
densities \cite{JL}: a high density region which can be thought of as
a traffic jam behind the defect and a low density region in front of
the defect.  Here the presence of the drive appears necessary for the
defect to induce the phase separation.  Moving defects ({\it i.e.}
particles with dynamics different from that of the others) have also
been considered and exact solutions obtained
\cite{DJLS,Mallick,DE99,Sasamoto3}.  For the simple case of a slowly
moving particle the phenomenon of a queue of particles forming behind
it has been shown to be analogous to Bose condensation \cite{Evans96,KF}.

\section{Stochastic Ballistic Annihilation and Coalescence}
\label{Sec:SBAC}

In Section~\ref{Sec:SBACintro} we introduced two reaction systems,
namely annihilation ($A+A \to \emptyset$) and coalescence ($A+A \to
A$).  As noted in Section~\ref{Sec:SBACintro}, it is well known that
annihilation and coalescence reactions are equivalent when the
reactant motion is diffusive \cite{Spouge,Peliti} and hence lead to
the same $t^{-1/2}$ density decay in one dimension.

We now consider in detail the distinct case of ballistic reactant
motion by which it is meant that particles move with some fixed
velocity.  We study a class of models that comprise an arbitrary
combination of annihilation and coalescence of particles with two
(conserved) velocities and stochastic reaction dynamics and which was
solved in \cite{BEK}.  As we shall see, an exact solution is possible
by virtue of a matrix product method in which the matrices involved
can be written in terms of the raising and lowering operators of the
$q$-deformed harmonic oscillator.  A related, but distinct, ballistic
reaction model (not contained within the class discussed here) had
also revealed a connection between ballistic reaction systems and
$q$-deformed operator algebras \cite{Richardson}.  

\begin{figure}[htb]
\center{\includegraphics[scale=0.6]{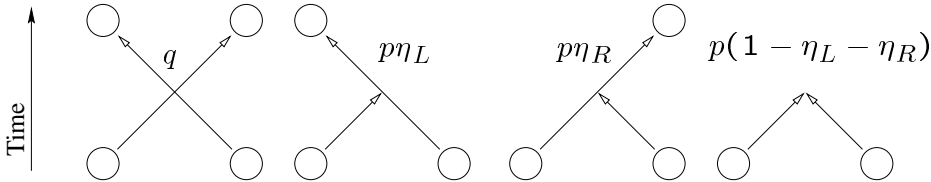}}
\caption{\label{fig:outcomes}The possible reactions and their probabilities.}
\end{figure}

We now define the class of models to be considered.  At time $t=0$
reactants are placed randomly on a line.  Each particle is assigned a
velocity $+c$ (right-moving) or $-c$ (left-moving) with probability
$f_R$ and $f_L = 1 - f_R$ respectively.  Particles move ballistically
until two collide, at which point one of four outcomes follows, see
Figure~\ref{fig:outcomes}: the particles pass through each other with
probability $q$; the particles coalesce into a left (right) moving
particle with probability $p \eta_L$ ($p \eta_R$); the particles
annihilate with probability $p(1-\eta_L-\eta_R)$.  Here $p=1-q$ is the
probability that some reaction occurs.

\begin{figure}[b]
\center{\includegraphics[scale=0.8]{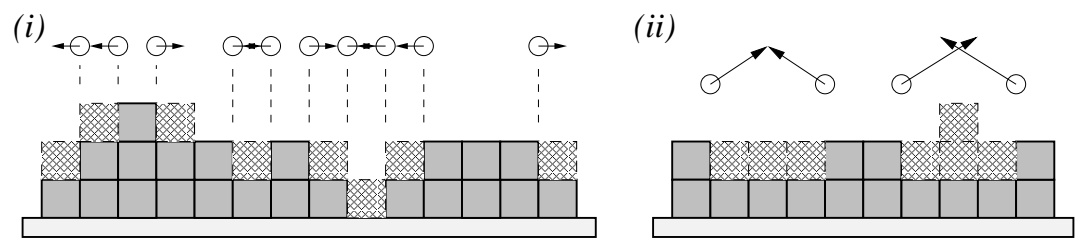}}
\caption{\label{fig:RSOS}(i) Mapping of a surface growth model to a
particle reaction system through the identification of upward and
downward steps to left- and right-moving particles.  Addition of new
particles that stick only to the sides of the surface cause the
particles to move.  (ii) The correspondence of particle annihilation
and scattering to the elimination and nucleation of a terrace
respectively.}
\end{figure}

As an example of an application of this model consider the
identification of right- and left-moving particles with the edges of
terraces as illustrated in Figure~\ref{fig:RSOS}.  If new particles
are added to the system in such a way that they only stick to the
sides, and the rate of particle addition is taken to infinity, one
obtains ballistic motion of terrace edges.  When two edges meet, a
terrace is completed which corresponds to annihilation of approaching
particles.  Hence the relevant parameters of the general annihilation
coalescence model are $\eta_L = \eta_R =0$.  The  scattering
reaction, which occurs with rate $q$, corresponds to the possibility of
a new terrace being formed when two edges meet.

The case  of deterministic
annihilation $q=\eta_L = \eta_R =0$ 
was studied by Elskens and Frisch \cite{EF}.  Here particles
always annihilate on contact.  This case was found to exhibit a
density decay that depends on time as $t^{-1/2}$, but only if the
initial densities of the two particle species (left- and right-moving)
are the same.  In the following analysis of the more general class of
reaction systems, we find that such a result persists if two
`effective' initial densities (to be defined below) are equal.
Furthermore, we will see that the introduction of stochasticity into
the reaction dynamics {\it i.e.} the parameter $q$, induces a transition
in the density decay form.

\begin{figure}
\center{\includegraphics[scale=0.6]{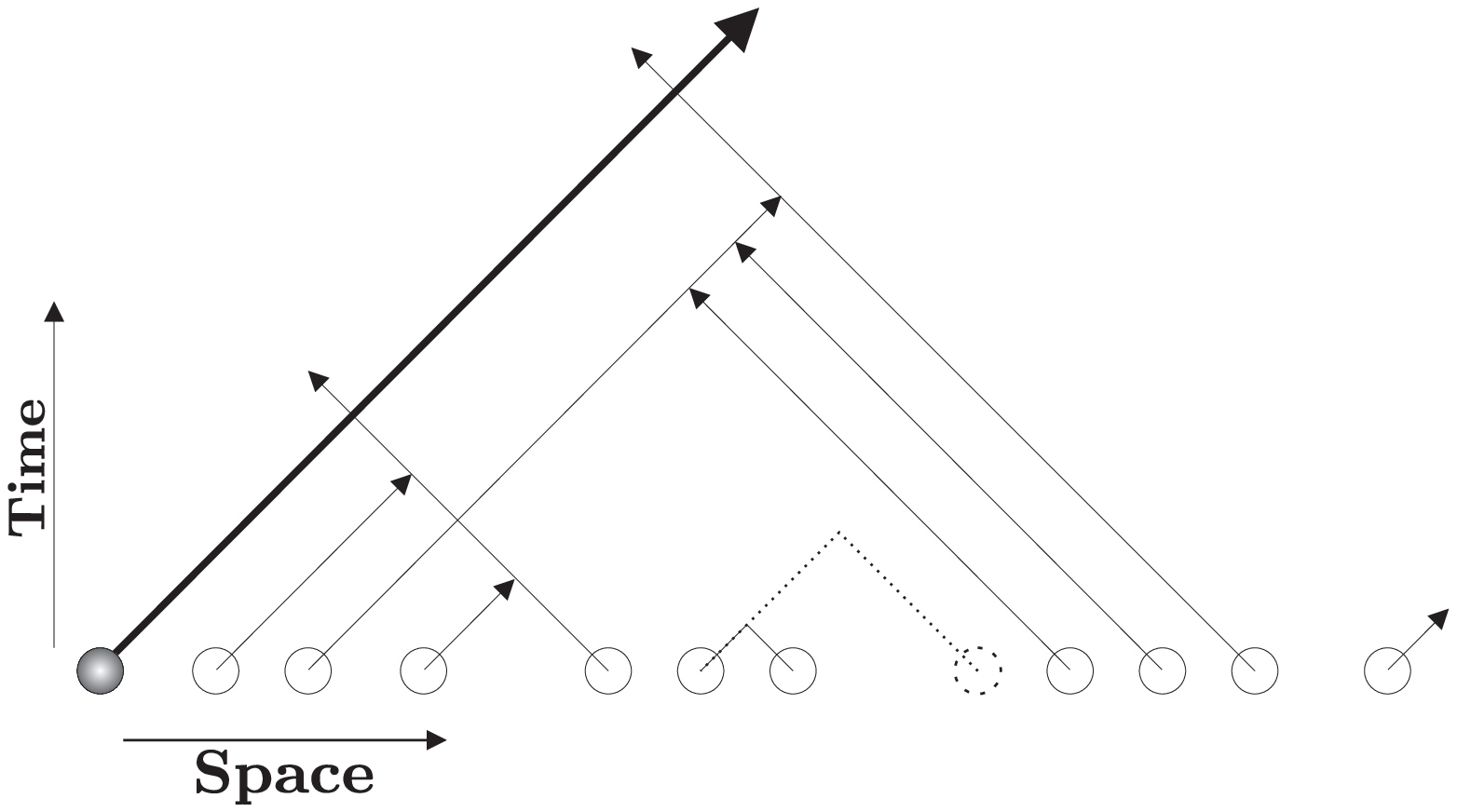}}
\caption{\label{fig:traj} 
  A configuration and set of trajectories and reactions for a
  test particle (shaded and bold line) encountering a string of $N=10$
  particles.  Note how changing the spacing between, for example, the
  fifth and sixth particles (indicated by dotted lines) alters the
  time sequence of the reactions but not the final survival
  probability.}
\end{figure}
Our aim is to calculate the density decay. To do this we shall
consider without loss of generality a right moving test particle as
illustrated in figure~\ref{fig:traj}.  We define $P_S^{(R)}(t)$ as the
probability that the test particle survives up to a time $t$
respectively.  From figure~\ref{fig:traj} one can see that the initial
spacing of the particles on the line does not affect the {\em
sequence} of possible reactions for any given particle, in particular
for the test particle (we return to this point later).  Also note that
after a given time $t$, the test particle may only have interacted
with the $N$ particles initially placed within a distance $X=2ct$ (and
to the right) of the chosen particle.  These two facts imply that the
survival probability can be expressed in terms of two independent
functions.  The first is $G(N;X)$, the probability that initially
there were exactly $N$ particles in a region of size $X=2ct$.  The
second is $F(N)$, the probability that the test particle survives
reactions with the $N$ particles initially to its right, and depends
only on the sequence of the $N$ particles. Explicitly,
\begin{equation}
\label{eqn:Psum}
P_S^{(R)}(X) = \sum_{N=0}^{\infty} G(N;X) F(N)\;.
\end{equation}

Thus the problem is reduced to two separate combinatorial problems of
calculating $G(N;X)$ and $F(N)$.  Note that the choice of $G(N;X)$
allows one to consider a model defined on a lattice or on the real
line.  In the present work we assume particles are initially placed on
a line with nearest-neighbour distances chosen independently from an
exponential distribution with unit mean.  Then we can use a well-known
result \cite{Feller}
\begin{equation}
G(N;X) = X^N \mbox{e}^{-X}/N!\;.
\end{equation}   
We now show that the second problem (calculation of
$F(N)$) can be formulated within a matrix
product approach.

As an example consider a test particle encountering the string of
reactants depicted in Figure~\ref{fig:traj}. We claim that the
probability of the test particle surviving through this string may be
written as
\begin{equation}
\langle W | RRRLRLLLLR | V \rangle
\label{eqn:Fex}
\end{equation}
where $R,L$ are matrices  and $\langle W |$, $| V
\rangle$ vectors with scalar product $\langle W | V \rangle = 1$.
Thus we write, in order, a matrix $R$ ($L$) for each right (left)
moving particle in the initial string.

The conditions for an expression such as (\ref{eqn:Fex}) to hold for
an arbitrary string are, in fact, rather intuitive
\begin{eqnarray}
\label{eqn:RL}
RL &=& qLR + p(\eta_L L + \eta_R R + [1-\eta_L -\eta_R]) \\
\label{eqn:Wdef} \langle W| L &=&  \langle W| (q+p\eta_R) \\
R |V\rangle &=& |V\rangle\;. \label{eqn:Vdef}
\end{eqnarray}
The condition (\ref{eqn:RL}) just echoes the reactions that occur in
Figure~\ref{fig:outcomes} {\it i.e.}  after an interaction between a
right-moving and left-moving particle there are four possible outcomes
(see Figure~\ref{fig:outcomes}) corresponding to the four terms on the
right hand side of (\ref{eqn:RL}) with probabilities given by the
respective coefficients. Using (\ref{eqn:RL}), any initial matrix
product such as (\ref{eqn:Fex}) can be reduced to a sum of terms of
the form $\langle W | L^sR^t | V \rangle$ corresponding to all possible
final states ensuing from the initial string and with coefficients
equal to the probabilities of each final state.  These final states
give the possible sequences of particles that the right-moving test
particle will encounter.  The test particle will survive such a final
state and pass through the $s$ left-moving particles with probability
$(q+p\eta_R)^s$. The condition (\ref{eqn:Vdef}) ensures
that this probability is obtained for each possible final state {\it i.e.}\
\begin{equation}
\langle W | L^sR^t | V \rangle = (q+p\eta_R)^s \langle W | R^t | V \rangle\;.
\label{left}
\end{equation}
Finally the condition (\ref{eqn:Wdef}), along with $\langle W | V
\rangle = 1$, ensures that once a right-moving particle passes through
all the left-moving particles in the string it no longer plays a role.
Thus (\ref{left}) becomes
\begin{equation}
\langle W | L^sR^t | V \rangle = (q+p\eta_R)^s \;.
\end{equation}

Note that the reason for a matrix product formulation is different
from the PASEP. There the steady state probability of any
configuration, for arbitrary system size, could be written as a matrix
product.  Here the probability of a particle surviving some arbitrary
sequence of particles can be written as a matrix product.

The above approach relies on an important property of the system which
is invariance of a reaction sequence with respect to changes of
initial particle spacings.  To understand this, consider again
Figure~\ref{fig:traj}.  By altering the initial spacings of the
particles, the absolute times at which trajectories intersect and
reactions may occur (if the reactants have survived) may be altered.
For example, by increasing the spacing between the fifth and sixth
particles, the trajectories of the third and fourth particles can be
made to intersect first. However as we have already seen, for any
particle, the {\em order} of intersections it encounters does not
change and so the final states and probabilities are invariant. This
invariance is manifested in the matrix product by the fact that the
order in which we use the reduction rule (\ref{eqn:RL}) is unimportant
{\it i.e.}\ matrix multiplication is associative.  Thus, it is the
invariance with respect to initial spacings that allows the system to
be solved by using a product of matrices.

We now proceed to evaluate $F(N)$. 
 Averaging (\ref{eqn:Fex}) over all
initial strings of length $N$, recalling
that $f_R$ and $f_L$ are the probabilities that a particle is assigned velocity
$+c, -c$ respectively, yields
\begin{equation}
F(N) = \langle W | (f_L L + f_R R)^N | V \rangle\;.
\label{eqn:F}
\end{equation}
To evaluate $F(N)$ we first write 
\begin{eqnarray}
R = \frac{\sqrt{f_R^\ast f_L^\ast}}{f_R} \, \hat{a} +
\eta_L\quad\mbox{and}\quad
L = \frac{\sqrt{f_R^\ast f_L^\ast}}{f_L} \, \hat{a}^\dagger + \eta_R \;.
\end{eqnarray}
In these equations we have introduced two new parameters
\begin{equation}
f_R^\ast = f_R (1 - \eta_L) \quad \mbox{and} \quad
f_L^\ast = f_L (1 - \eta_R)
\end{equation}
which we call effective initial particle densities and whose ratio
\begin{equation}
\chi = \frac{f_R^\ast}{f_L^\ast}
\end{equation}
turns out to be an important parameter of the model.

One can verify from (\ref{eqn:RL}--\ref{eqn:Vdef}) that $\hat{a},
\hat{a}^\dagger$ satisfy a $q$-deformed harmonic oscillator algebra
\begin{eqnarray}
\hat{a}\hat{a}^\dagger - q \hat{a}^\dagger \hat{a} &=& 1-q
\label{eqn:ho} \\
  \langle W| a^\dagger &=&  \langle W|q/\sqrt{\chi} \\
  a |V\rangle &=& \sqrt{\chi}\, |V\rangle\;.
\label{eqn:coherent}
\end{eqnarray}
Thus (\ref{eqn:F}) can be written as 
\begin{equation}
F(N) = \langle W | \left[ \sqrt{f_R^\ast f_L^\ast}\, (\hat{a} +
 \hat{a}^\dagger) + 1{-}f_L^\ast{-}f_R^\ast \right]^N | V \rangle\;.
\label{eqn:Fmat}
\end{equation}
The vectors $\bra{W}$ and $\ket{V}$ are eigenvectors of
$\hat{a}^\dagger,\hat{a}$, and as in Section~\ref{Sec:PASEP}, they can be
explicitly calculated.

At this point one can see that we have precisely the same mathematical
structure as we did when solving for the steady state of the
PASEP---compare (\ref{eqn:ho}--\ref{eqn:coherent}) with
(\ref{pasepqho1}--\ref{pasepqho3}).  We just need to carry out the
same procedure of diagonalising a matrix that is linear combination of
the identity operator and the position operator of a $q$-deformed
harmonic oscillator.  Rather than repeat the details here we  go
directly to a discussion of the phase diagram.


\begin{exercise}{Deterministic ballistic annihilation and random walks}

The Elskens-Frisch model \cite{EF} of deterministic ballistic annihilation is a
special case of the stochastic model described above that has the
parameters $q = \eta_L = \eta_R = 0$.  For the case where particles
are initially placed on all sites of a lattice with equal probability
of right- and left-moving particles, write down all possible initial
conditions that would allow a right-moving test particle to survive
through one, two and three sites.  Hence relate the survival
probability of the test particle after $N$ sites to the probability of
a random walker not returning to the origin after $N+1$ steps.  Hint:
devise a criterion for the survival of the test particle that involves
the relative number of left- and right-moving particles for a given
initial condition.

Construct the matrices $R$ and $L$ and the vectors $\bra{W}$ and
$\ket{V}$ and explain how the resulting expression for $F(N)$ (with
$f_L=f_R=\frac{1}{2}$) is related to the random walk problem.

\end{exercise}


\subsection{Phase diagram}
We consider the long-time density decay
$\varrho(t)$
which is obtained from the survival probabilities
$P_S^{(R)}(t)$, $P_S^{(L)}(t)$
for right and left moving test particles  through
\begin{equation}
\label{eqn:rhodef}
  \varrho(t) = f_R  P_S^{(R)}(t) + f_L P_S^{(L)}(t)\;.
\end{equation}
($P_S^{(L)}(t)$ can be deduced from
$P_S^{(R)}(t)$ through the left-right symmetry of the model.)

\begin{figure}
\center{\includegraphics[scale=0.8]{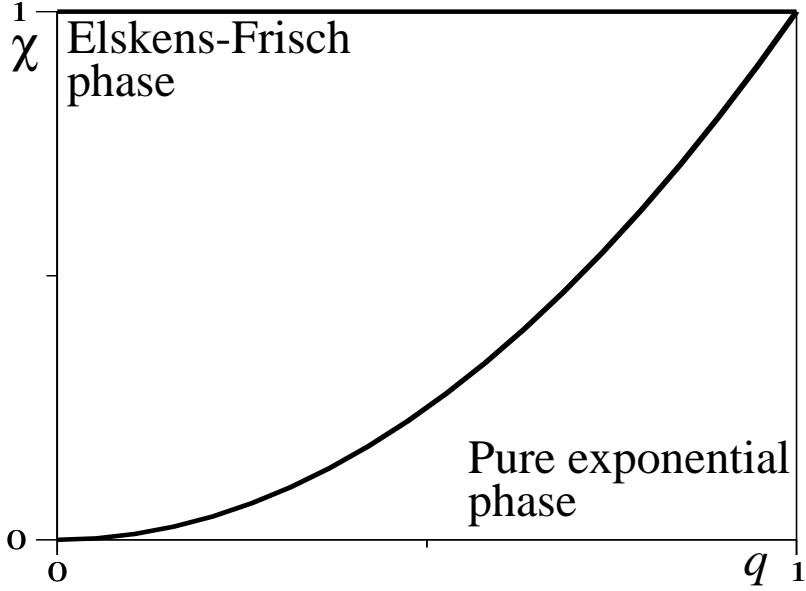}}
\caption{\label{fig:SBACpd} 
Phase diagram for the model of stochastic ballistic annihilation and
coalescence.  The density decay form depends on the ratio of effective
densities $\chi$ and the stochasticity parameter $q$.}
\end{figure}

The matrix product calculation for $F(N)$ implies that
the density decays as
\begin{equation}
\varrho(t) = \varrho_{\infty} + \Delta(t) \;.
\end{equation}
In this equation $\varrho_{\infty} = f_L (1 - \chi)$ is the fraction
of particles remaining once no more reactions are possible ({\it i.e.}\ all
particles moving in the same direction).
We now list the results the residual density
$\varrho_{\infty}$
when  $\chi < 1$.
(Results for $\chi>1$ one be deduced using the left-right symmetry of the
model.)

For $ \chi < q^2 $
$$\Delta(t) = \displaystyle f_R \left( 1 - \frac{f_R^\ast}{q^2 f_L^\ast} \right) 
\exp\left[
 - 2\, \left(1-q\right) \left( f_L^\ast - f_R^\ast/q \right) ct \right]$$
For $ \chi = q^2 $
$$\Delta(t) = \displaystyle \frac{f_R}{\sqrt{2\pi}}
 \left( \frac{1}{f_L^\ast f_R^\ast}
  \right)^{\!\frac{1}{4}} 
\frac{\exp\left[-2(\sqrt{f_L^\ast} - \sqrt{f_R^\ast})^2 c t\right]}
  {(ct)^{1/2}} $$ 
For $ q^2 {<} \chi {<} 1 $ 
\begin{multline}
$$\displaystyle \Delta(t) = \displaystyle
\frac{(q;q)_\infty^4}{4\sqrt{2\pi}(q\sqrt\chi,q/\sqrt\chi;q)_\infty^2}$$ \\
$$\displaystyle
\times \left( \frac{1}{f_L^\ast f_R^\ast} \right)^{\!\frac{3}{4}}
 \frac{f_L f_R^\ast + f_R f_L^\ast}{( \sqrt{f_L^\ast} -
   \sqrt{f_R^\ast})^2}
 \frac{ \exp\left[-2 (\sqrt{f_L^\ast} - \sqrt{f_R^\ast})^2 ct\right]}{(ct)^{3/2}}$$\nonumber
\end{multline}
For $ \chi = 1 $
$$\Delta(t) = 
\displaystyle \frac{1}{\sqrt{2\pi}} \left( \frac{1}{f_L^\ast f_R^\ast}
  \right)^{\!\frac{1}{4}} \left(\frac{1}{ct} \right)^{\!\frac{1}{2}}$$

We now consider the phase diagram---figure~\ref{fig:SBACpd}. This is
spanned by the values of $\chi$ (the ratio of effective densities
$\chi = f_R^\ast / f_L^\ast$) and $q$ (the stochasticity parameter):
$\chi$ parameterises all information concerning the asymmetry between
the right and left moving particles whereas $q$ parameterises the
level of stochasticity.  Thus there is universality of ballistic
annihilation and coalescence: for a generic choice of $\eta_R$,
$\eta_L$ defining a particular annihilation-coalescence model, the
same four decay regimes are found by varying the initial densities or
stochasticity parameter $q$.  In this way the universality can be
considered as a `law of corresponding states'.

The line $\chi =1$ (equal effective densities) is non-generic since a
single, power law, decay regime is found.  The decay does not depend
on the stochasticity $q$.  This special line was found by Elskens and
Frisch \cite{EF} for the case of deterministic annihilation
($\eta_L=\eta_R=q=0$) and we have shown that this special line is
present for all combinations of annihilation and coalescence.  This
phase can be understood through the picture of \cite{EF}.  Density
fluctuations in the initial conditions lead to trains of left- and
right-moving particles: in a length $\sim t$ the excess particle
number is $\sim t^{1/2}$ which yields the $t^{-1/2}$ density decay.
At long times, the train size is large and so a particle in one train
encounters many particles in the other and will eventually react
making the parameter $q$ irrelevant.

The second phase found in \cite{EF} (labelled the Elskens-Frisch phase
in the diagram) is seen to persist for nonzero $q$ and in this phase
the two particle species decay at equal rates.  The two new phases
that arise in the full model
for $\chi \le q^2$ ({\it i.e.}\ as a consequence of
randomness in the reaction dynamics) have the contrasting property
that two particle species decay at {\it unequal} rates leaving a
non-zero population of left-moving particles.  A simple example of
non-equal decays is the case $\eta_R=0, \eta_L=1$ ($\chi =0$).  Then
left-moving particles do not decay but simply absorb the right-moving
particles with probability $1-q$ giving $\varrho = f_R {\rm e}^{-2
  (1-q) f_L c t}$.  Our results show that, in general, increasing $q$
leads to a non-trivial transition at $\chi=q^2$ to a regime where the
two species have different decay forms.

\section{Conclusion}
In these lectures we have tried to give a flavour of the collective
phenomena exhibited by simple low-dimensional systems with stochastic
dynamics.  Specifically we have seen that the partially asymmetric
exclusion process exhibits a number of phase transitions (of both
first and second order) and long-range correlations even in one
dimension.  We have also shown how similar mathematical techniques can
be applied in the distinct context of a particle reaction system to
obtain an exact solution.

We hope that the reader with a background in equilibrium statistical
physics is pleasantly surprised by the diversity of phenomena that
emerge in low-dimensional nonequilibrium systems.  We have also
endeavoured to illustrate how novel analytical techniques are being
developed for these systems.  Through the inclusion of background
material and the exercises, we hope also to have inspired confidence
in the reader to approach the literature and make an active
contribution to this fertile field.

\noindent {\bf Acknowledgments}:
We would like to thank our collaborators 
F. Colaiori, F. Essler and Y. Kafri with whom
the work of section 5 and section 6 was carried out
and also M. J. E. Richardson who contributed to the 
early development of that work.
%



\end{document}